\documentclass[aps,nofootinbib,floatfix,amsmath,amssymb,preprint,groupedaddress,groupedaddress]{revtex4}

\usepackage{hyperref}

\usepackage{rotating}

\usepackage[font=small,labelfont=bf]{caption}
\usepackage{subfig}

\usepackage{booktabs}
\usepackage{dcolumn}
\usepackage{array}
\usepackage[english]{babel}

\usepackage{xspace}

\usepackage{bm}

\newcommand{\mxnewcommand}[2]{\newcommand{#1}{\ensuremath{#2}\xspace}}

\newcommand{\xnewcommand}[2]{\newcommand{#1}{#2\xspace}}

\mxnewcommand{\gev}{\,\text{GeV}}
\mxnewcommand{\tev}{\,\text{TeV}}
\mxnewcommand{\invfb}{/\text{fb}}

\newcommand{\roots}[1]{\ensuremath{\sqrt{s}={#1}\tev}}

\mxnewcommand{\mhalf}{m_{1/2}}
\mxnewcommand{\mzero}{m_0}
\mxnewcommand{\azero}{A_0}
\mxnewcommand{\tanb}{\tan\beta}
\mxnewcommand{\sgnmu}{\text{sign}\,\mu}
\mxnewcommand{\msbar}{\overline{\text{MS}}}
\mxnewcommand{\invalpha}{1/\alpha_{\text{em}}(M_Z)^{\msbar}}
\mxnewcommand{\alphas}{\alpha_s(M_Z)^{\msbar}}
\mxnewcommand{\mt}{m_t^\text{Pole}}
\mxnewcommand{\mb}{m_b(m_b)^{\msbar} }

\mxnewcommand{\mts}{m_t}
\mxnewcommand{\mbs}{m_b}
\mxnewcommand{\invalphas}{1/\alpha_{\text{em}} }
\mxnewcommand{\alphass}{\alpha_s}

\renewcommand{\(}{\left(}
\renewcommand{\)}{\right)}
\renewcommand{\[}{\left[}
\renewcommand{\]}{\right]}

\mxnewcommand{\like}{\mathcal{L}}
\mxnewcommand{\prior}{\pi}
\mxnewcommand{\params}{m}
\mxnewcommand{\ev}{\mathcal{Z}}
\mxnewcommand{\point}{\vec{x}}
\mxnewcommand{\model}{\text{model}}
\mxnewcommand{\data}{\text{data}}
\mxnewcommand{\pd}{\,\prod\text{d}}
\newcommand{\priorf}[1]{\ensuremath{\prior(#1)}}
\newcommand{\likef}[1]{\ensuremath{\like(#1)}}
\newcommand{\gaussian}[3]{\ensuremath{\exp \[-\frac{\(#1 - #2\)^2}{2#3} \]}}
\newcommand{\pg}[2]{\ensuremath{p(#1\,\bm{|}\,#2)}}
\newcommand{\p}[1]{\ensuremath{p(#1)}}

\newcommand{\s}[1]{\ensuremath{\tilde{#1}}}
\newcommand{\neut}[1]{\ensuremath{{\chi}^0_{#1}}}

\newcommand{\ms}[1]{\ensuremath{m_{\s{#1}}}}
\newcommand{\mneut}[1]{\ensuremath{m_{\neut{#1}}}}

\mxnewcommand{\sigsip}{\sigma^{\text{SI}}_p}
\mxnewcommand{\abund}{\Omega h^2}
\newcommand{\br}[1]{\ensuremath{\text{BR}}(#1)}
\mxnewcommand{\bsg}{\br{B_s\to X_s\gamma}}
\mxnewcommand{\bsmm}{\br{B_s\to \mu\mu}}
\mxnewcommand{\btn}{\br{B_u\to \tau\nu}/\br{B_u\to \tau\nu}|_{\text{SM}}}
\mxnewcommand{\damu}{\delta a_\mu}
\mxnewcommand{\mh}{m_h}
\mxnewcommand{\ma}{m_A}
\mxnewcommand{\mw}{M_W}
\mxnewcommand{\dmbs}{\Delta M_{B_s}}
\mxnewcommand{\sineff}{\sin^{2} \theta_{\ell,\text{eff}}}

\mxnewcommand{\pmm}{(\mzero,\,\mhalf)}
\mxnewcommand{\pat}{(\azero,\,\tanb)}
\mxnewcommand{\pcs}{(\mneut{1},\,\sigsip)}
\xnewcommand{\stauc}{stau-coannihilation}
\xnewcommand{\Stauc}{Stau-coannihilation}

\let\oldcite\cite
\renewcommand{\cite}{~\oldcite}
\newcommand{\reftable}[1]{Table~\ref{#1}} 
\newcommand{\reffig}[1]{Fig.~\ref{#1}}       
 
\newcommand{\refeq}[1]{Eq.~(\ref{#1})}
\newcommand{\refsec}[1]{Sec.~\ref{#1}}
\newcommand{\refcite}[1]{Ref.\cite{#1}}            
\xnewcommand{\eg}{\textit{e.g.,}}
\xnewcommand{\ie}{\textit{i.e.,}}
\mxnewcommand{\dash}{\text{--}}

\newcommand{\beq}{\begin{equation}}
 \newcommand{\eeq}{ \end{equation}}

\begin{document}

\title{Prospects for constrained supersymmetry at \roots{33} and \roots{100} proton-proton super-colliders}
  
\author{Andrew Fowlie}
\email{Andrew.Fowlie@KBFI.ee}
\affiliation{National Institute of Chemical Physics and Biophysics, Ravala 10,
Tallinn 10143, Estonia}

\author{Martti Raidal}
\email{Martti.Raidal@CERN.ch}
\affiliation{National Institute of Chemical Physics and Biophysics, Ravala 10,
Tallinn 10143, Estonia}

\date{\today}

\begin{abstract}
Discussions are under way for a high-energy proton-proton collider.  Two preliminary ideas are the \roots{33} HE-LHC and the \roots{100} VLHC.  With Bayesian statistics, we calculate the probabilities that the LHC, HE-LHC and VLHC discover SUSY in the future, assuming that nature is described by the CMSSM and given the experimental data from the LHC, LUX and Planck. We find that the LHC with $300\invfb$ at \roots{14} has a $15\dash75\%$ probability of discovering SUSY.  Should that run fail to discover SUSY, the probability of discovering SUSY with  $3000\invfb$ is merely $1\dash10\%$. Were SUSY to remain undetected at the LHC, the HE-LHC would have a $35\dash85\%$ probability of discovering SUSY with $3000\invfb$. The VLHC, on the other hand, ought to be definitive;  the probability of it discovering SUSY, assuming that the CMSSM is the correct model, is $100\%$. 
\end{abstract}

\maketitle

\section{\label{sec:intro}Introduction} 

Supersymmetry\cite{Salam:1974yz,Haber:1984rc,Nilles:1983ge} (SUSY) is the most popular scenario of new physics beyond the Standard Model (SM).
It stabilizes the electroweak scale against radiative corrections from any high-scale physics such as a Grand Unified Theory\cite{Georgi:1974sy,Dimopoulos:1981zb} (GUT),
predicts gauge coupling unification, and provides a natural framework for explaining the observed amount of dark matter (DM)
via a thermal relic density of the lightest supersymmetric particle\cite{Jungman:1995df}.  Despite huge experimental effort in the last few years, 
no clear experimental evidence for the existence of
SUSY has been found so far in collider experiments, direct and indirect searches for DM or in precision physics. 
The Large Hadron Collider (LHC) upgrade with a center-of-mass energy of \roots{14} and proposed future colliders, such as the High-Energy LHC (HE-LHC)\cite{Assmann:1284326} with \roots{33} or 
the Very Large Hadron Collider (VLHC) with \roots{100} (see \eg \refcite{FCC}) could, therefore, play a crucial role in deciding the fate of SUSY. 

Direct hadron collider searches for SUSY and indirect searches for DM are complementary; hadron collider searches are predominantly sensitive to the masses and mass hierarchies of the colored sparticles, whereas indirect searches for DM are sensitive to the mass and character (\ie gaugino and higgsino admixture) of the neutralino.  We include LHC and indirect searches in our analysis.

The aim of this paper is to quantify with Bayesian statistics the probability of discovering the Constrained Minimal Supersymmetric Standard Model (CMSSM) 
at the \roots{14} LHC, HE-LHC and VLHC, given that previous experiments found no discrepancies 
with the predictions of the SM. We choose to work with the CMSSM because it is the best-known SUSY model with non-degenerate
mass spectra. 
We believe that our results help to motivate building the HE-LHC and the VLHC, and contribute to forming research programs for those colliders. 

Given our assumptions, we find that there is an appreciable probability that the LHC or HE-LHC could discover the CMSSM. 
However, the VLHC would be definitive --- either the VLHC discovers the CMSSM or one has to relax
some of the assumptions we have made. This conclusion results from the fact that the CMSSM mass spectrum cannot be arbitrarily heavy,
else the Higgs mass would be greater than $\sim125\gev$ and the DM abundance could not be explained
with thermal freeze-out processes.
Since those are physical requirements, a similar conclusion must be obtained for other SUSY models
with similar numbers of free parameters. 
Our results do not apply to SUSY models with compressed mass spectra\cite{LeCompte:2011cn}, which could evade searches for missing transverse momentum, or to models with extended particle content, such as the NMSSM\cite{Fayet:1974pd,Ellwanger:2009dp}, that have additional tree-level contributions to the 
Higgs boson mass as well as new candidates for DM. Those models deserve separate dedicated studies.

The paper is organized as follows. In the next section we present the basics of Bayesian statistics on which our methodology is based.
In \refsec{sec:posterior}, we calculate the posterior pdf and in \refsec{sec:accessible}, we discuss which parts of the CMSSM parameter
space might be discoverable at the LHC \roots{14}, a \roots{100} VLHC or a \roots{33} HE-LHC and 
complete our calculation, finding the probability that a collider could find 
constrained supersymmetry.

\section{\label{Bayesian}Bayesian approach}

To quantify statements such as, ``the probability that the HE-LHC could discover constrained supersymmetry,'' we invoke Bayesian statistics.  We remind the reader that in Bayesian statistics, probability is a numerical
measure of belief in an hypothesis. See \eg \refcite{Trotta:2008qt} for a pedagogical introduction.

From an experiment, one can construct a ``likelihood function,'' describing the probability of obtaining the data, given a particular point, \point, in a model's parameter space,
\beq
\likef{\point} = \pg{\data}{\point,\model}.
\eeq
The likelihood function for a measurement is typically a Gaussian function (by the central limit theorem). Note that the likelihood function, 
however, is not a probability distribution function (pdf). Bayes' theorem,
\beq
\pg{a}{b} = \frac{\pg{b}{a}\p{a}}{\p{b}},
\eeq
permits us to ``invert''  the likelihood function to find the probability density of the model's parameter space, given the data;
\beq
\label{eq:bayes}
\pg{\point}{\data,\model} = \frac{\pg{\data}{\point,\model}\pg{\point}{\model}}{\pg{\data}{\model}}.
\eeq
This quantity, which we shall call the ``posterior,'' is central to our work; it is a numerical measure of our belief in the parameter space \textit{after} seeing the experimental data. 
On the other hand, $\priorf{\point} \equiv \pg{\point}{\model}$, the ``prior,'' is a numerical measure of our
belief in the parameter space \textit{before} seeing the experimental data. For our purposes, the denominator is merely a normalization factor. One can see that the likelihood function ``updates'' our prior
beliefs with experimental data, resulting in our posterior beliefs.

The posterior is a pdf of the continuous model parameters \point. As such, if we wish to find the probability within a region of parameter space, 
we integrate (in the lexicon of Bayesian statistics, ``marginalize''), \eg
\beq
\pg{\point \in A}{\data,\model} = \int\limits_{A} \pg{\point}{\data,\model} \pd x.
\eeq

In our analysis, the model is the Constrained Minimal Supersymmetric Standard Model (CMSSM)\cite{Chamseddine:1982jx,Arnowitt:1992aq,Kane:1993td}, the data is that from, \textit{inter alia,} the LHC, LUX and Planck, and \point is a parameter point
in the CMSSM.  The CMSSM has four continuous parameters, that is three soft-breaking parameters (\mhalf, \mzero and \azero) and the ratio of the Higgs vevs (\tanb),  and one discrete parameter, the sign of the Higgs parameter in the superpotential (\sgnmu).

We assume that particular regions of our model's parameter space would be discoverable at the \eg VLHC, whereas the complement of those regions would be 
entirely inaccessible;
\beq
\pg{\text{Discoverable at VLHC}}{\point,\model} = 
\left\{
\begin{array}{ll}
1  & \text{if } \point \in \text{Discoverable,} \\
0  & \text{if } \point \not\in \text{Discoverable}.
\label{eq:access}
\end{array}
\right.
\eeq
We find the probability that a point is discoverable by marginalizing;
\beq
\label{eq:pa}
\pg{\text{Discoverable}}{\data,\model} = \int \pg{\text{Discoverable}}{\point,\model}\pg{\point}{\data,\model} \pd x,
\eeq
for which we rely on the fact that
\beq
\pg{\text{Discoverable}}{\data,\point,\model}  = \pg{\text{Discoverable}}{\point,\model} ,
\eeq
\ie whether a point  in a model's parameter space is discoverable is dependent on only the point itself, and not on any previous measurements. 
Combining our \refeq{eq:access} with \refeq{eq:pa}, we find our desired result,
\beq
\label{eq:pfinal}
\pg{\text{Discoverable}}{\data,\model} = \int\limits_\text{Discoverable} \pg{\point}{\data,\model} \pd x.
\eeq
It can be shown that
\beq
\pg{\text{Discoverable, \model}}{\data} \approx \pg{\text{Discoverable}}{\data,\model}\priorf{\model},
\eeq
but it is an approximation that is reasonable only if no particular models are favored by the experimental data, \ie one should not, in general, multiply 
probabilities from \refeq{eq:pfinal} by his prior for the model in question; the normalization of the result will be incorrect.

\section{\label{sec:posterior}Posterior maps of the CMSSM}
We wish to calculate the posterior density of the CMSSM, given the experimental data in \reftable{tab:data}, which includes the Higgs mass, dark matter constraints, which we
assume to be the lightest neutralino,  $b$-physics observables,  electroweak precision observables and the anomalous magnetic moment of the muon
(see \eg
\refcite{Beskidt:2014oea,Cohen:2013kna,Henrot-Versille:2013yma,Buchmueller:2013rsa,Bechtle:2013mda,Kowalska:2013hha,Fowlie:2012im,Roszkowski:2012uf,Strege:2012bt,Allanach:2011wi,Akula:2012kk,Cabrera:2012vu}
for similar analyses).
The two ingredients that we require are the likelihood function
and the priors in \refeq{eq:bayes}. We will supply these ingredients to the nested sampling algorithm implemented in \texttt{MultiNest-2.18}\cite{Feroz:2008xx} via \texttt{PyMultiNest}\cite{Buchner:2014nha}, which will return the posterior. See \eg \refcite{Fowlie:2011mb} for a detailed introduction to the methodology.
\texttt{micrOMEGAS-2.4.5}\cite{Belanger:2010gh,Belanger:2006is} and \texttt{FeynHiggs-2.9.4}\cite{Heinemeyer:1998yj,Heinemeyer:1998np,Degrassi:2002fi,Frank:2006yh}
We construct Gaussian likelihoods for these experiments,
including  theory errors in quadrature. Our likelihood function is thus the product of these Gaussian functions,
\beq
\likef{\point} = \prod \gaussian{p_i(\point)}{\mu_i}{(\sigma_i^2+\tau_i^2)},
\eeq
where $p(\point)$ is our model's prediction at parameter point \point, $\mu$, $\sigma$ and $\tau$ are the mean, experimental error and theory error, respectively, and the product is taken over the data 
in \reftable{tab:data}. We calculate the CMSSM's predictions with \texttt{SOFTSUSY-3.3.7}\cite{Allanach:2001kg}, \texttt{micrOMEGAS-2.4.5}\cite{Belanger:2010gh,Belanger:2006is} and \texttt{FeynHiggs-2.9.4}\cite{Heinemeyer:1998yj,Heinemeyer:1998np,Degrassi:2002fi,Frank:2006yh}. Furthermore, we veto CMSSM points excluded at $95\%$ by LHC direct searches. We apply the $95\%$ exclusion contour from the ATLAS search in $20.1\invfb$ at \roots{8}\cite{ATLAS-CONF-2013-047} as a hard-cut on the \pmm plane. Although \refcite{ATLAS-CONF-2013-047} assumed that $\tanb=30$, $\azero=-2\mzero$ and $\mu>0$, \refcite{Allanach:2011ut,Bechtle:2011dm,Fowlie:2012im} demonstrated that the $95\%$ exclusion contour on the \pmm plane is independent of \tanb, \azero and \sgnmu. We implement the LUX direct search for dark matter\cite{Akerib:2013tjd} with an exclusion contour on the \pcs plane, however; the CMSSM's prediction for \sigsip contains a factor of $10$ uncertainty\cite{Ellis:2008hf}, which we incorporate with a method identical to that in \eg \refcite{Fowlie:2011mb}.

\begin{table}[t]
\begin{center}
\begin{ruledtabular}
\begin{tabular}{ccc}
Quantity & Experimental data, $\mu\pm\sigma$ & Theory error, $\tau$\\
\hline
\abund & $0.1199\pm0.0027$\cite{Ade:2013zuv} & $10\%$\cite{Allanach:2004jy,Allanach:2004jh}\\
\mh & $125.9\pm0.4\gev$\cite{Beringer:1900zz,Chatrchyan:2012ufa,Aad:2012tfa} & $2.0\gev$\cite{Allanach:2004rh} \\
\damu & $(28.8\pm7.9) \times10^{-10}$\cite{Beringer:1900zz} &  $1.0\times10^{-10}$\cite{Heinemeyer:2004gx}\\
\mw & $80.399\pm0.023\gev$\cite{Beringer:1900zz} & $0.015\gev$\cite{Heinemeyer:2004gx} \\
\sineff & $0.23116\pm0.00013\gev$\cite{Beringer:1900zz} & $0.00015\gev$\cite{Heinemeyer:2004gx} \\
\dmbs & $17.77\pm0.12\gev$\cite{Beringer:1900zz} & $2.4\gev$\cite{Trotta:2008bp} \\
\bsmm & $(3.2\pm1.5)\times10^{-9}$\cite{Beringer:1900zz} & $14\%$\cite{deAustri:2006pe}\\
\bsg & $(3.43\pm0.22)\times10^{-4}$\cite{Amhis:2012bh} & $0.21\times10^{-4}$\cite{Misiak:2006zs}\\
\btn & $1.43\pm0.43$\cite{Buchmueller:2011sw}\\
\hline 
\multicolumn{2}{l}{ATLAS $20.1\invfb$ at \roots{8}\cite{ATLAS-CONF-2013-047}}\\
\multicolumn{2}{l}{LUX $85.3$ live-days\cite{Akerib:2013tjd} with a factor of $10$ uncertainty in \sigsip\cite{Ellis:2008hf}}\\
\end{tabular}
\end{ruledtabular}
\end{center}
\caption{\label{tab:data}Experimental data included in our likelihood function. Note that in the case of \btn, the error listed is a combined theory and experimental error
from \refcite{Buchmueller:2011sw}.}
\end{table}

Because we are, \textit{a priori,} ignorant of the supersymmetry breaking scale, our priors for the soft parameters are invariant under rescalings, \ie we pick 
logarithmic priors for the soft masses \mzero and \mhalf,
\beq
\prior(x) \propto 1/x.
\eeq
We will, however, investigate linear priors by re-weighting our posterior. We pick linear priors for the remaining parameters \azero and \tanb. The principle motivation for supersymmetry is that it solves the ``fine-tuning'' problem that afflicts the SM\cite{Susskind:1978ms}, if the soft-breaking masses are sufficiently light\cite{Barbieri:1987fn,Ellis:1986yg}. For this reason, we restrict soft-breaking masses
in our priors to less than $20\tev$, in accordance with what one might have believed prior to seeing experimental data from \eg the LHC. Furthermore, \refcite{Kowalska:2013hha} indicates that this choice omits no credible parts of the parameter space. We include the SM nuisance parameters (\mts, \mbs, \alphass and \invalphas) with informative Gaussian priors, with their experimental means
and variances from the Particle Data Group (PDG)\cite{Beringer:1900zz}.  Our priors are listed in \reftable{tab:priors}. Furthermore, because our prior ranges and constraints are similar to those in \refcite{Kadastik:2011aa}, in which a simple ``hard-cut''  scanning algorithm was applied to the CMSSM, we can make a fair comparison between a Bayesian analysis and a simple method.

\begin{table}[t]
\begin{center}
\begin{ruledtabular}
\begin{tabular}{ccc}
Parameter & Distribution\\
\hline
\mzero & Log, $0.3\dash20\tev$\\
\mhalf & Log, $0.3\dash10\tev$\\
\azero & Flat, $|\azero|<5\mzero$\\
\tanb & Flat, $3\dash60$\\
\sgnmu & $\pm1$, with equal probability\\
\mb & Gaussian, $4.18\pm0.03\gev$\cite{Beringer:1900zz}\\
\mt& Gaussian, $173.07\pm0.89\gev$\cite{Beringer:1900zz}\\
\invalpha & Gaussian, $127.944\pm0.014$\cite{Beringer:1900zz}\\
\alphas & Gaussian, $0.1185\pm0.0005$\cite{Beringer:1900zz}\\
\end{tabular}
\end{ruledtabular}
\end{center}
\caption{\label{tab:priors}Priors for the CMSSM model parameters.}
\end{table}

\begin{figure}[t]
\centering
\subfloat[][Credible regions.]{\label{fig:m0m12:i}
\includegraphics[width=0.49\linewidth]{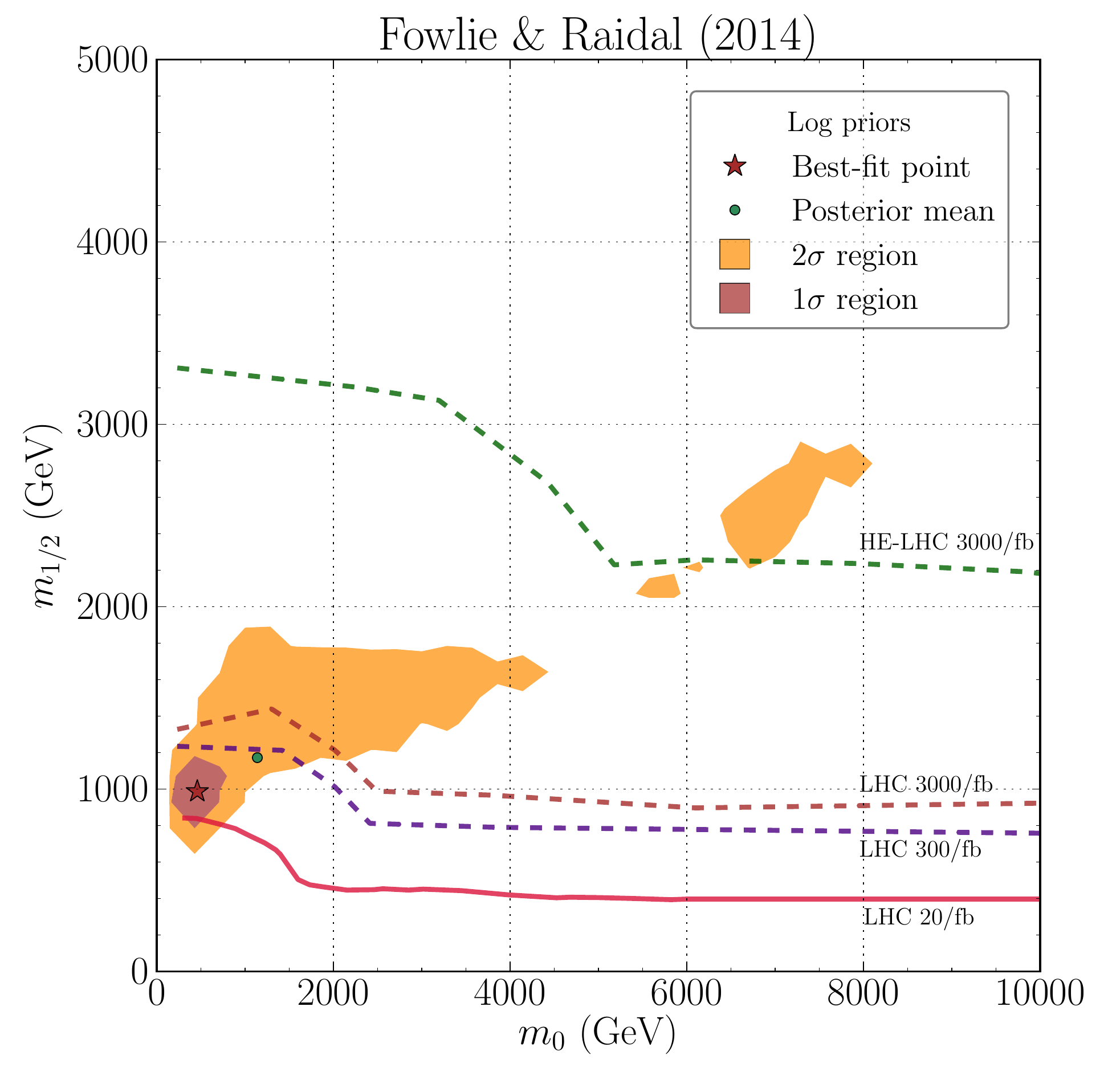}
}
\subfloat[][Scatter plot illustrating dark matter annihilation mechanisms. ]{\label{fig:m0m12:ii}
\includegraphics[width=0.49\linewidth]{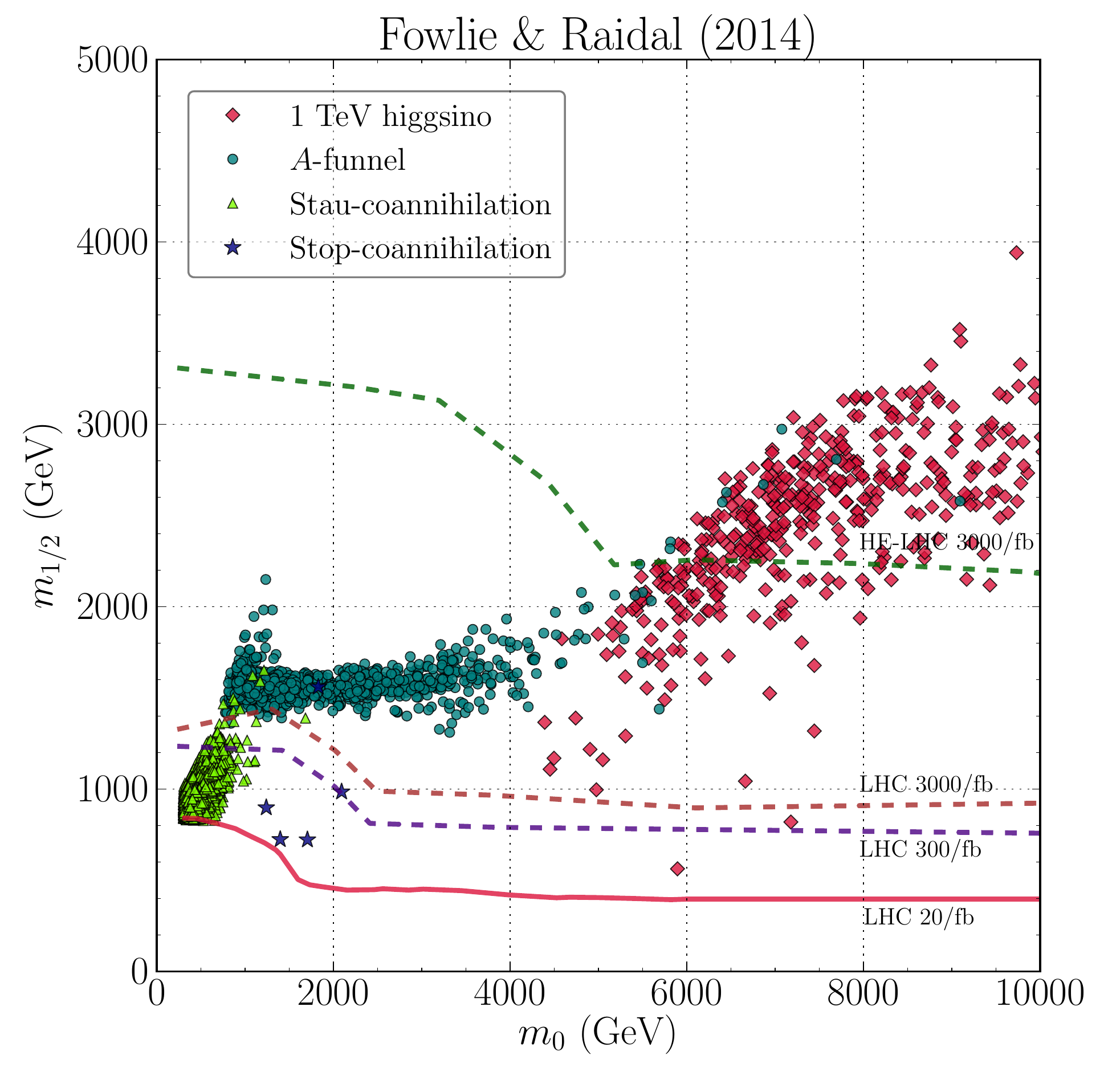}
}
\caption{The \pmm plane of the CMSSM showing: \protect\subref{fig:m0m12:i} The $68\%$ (red) and $95\%$ (orange) credible regions of the marginalized posterior and \protect\subref{fig:m0m12:ii}  A scatter of the points with appreciable posterior weight colored by their dominant dark matter annihilation mechanism. The expected discovery potentials of future hadron colliders are also shown.}
\label{fig:m0m12}
\end{figure}

\begin{figure}[t]
\centering
\subfloat[H][Credible regions on the \pmm plane with linear priors.]{\label{fig:m0m12_linear}%
\includegraphics[height=0.49\linewidth]{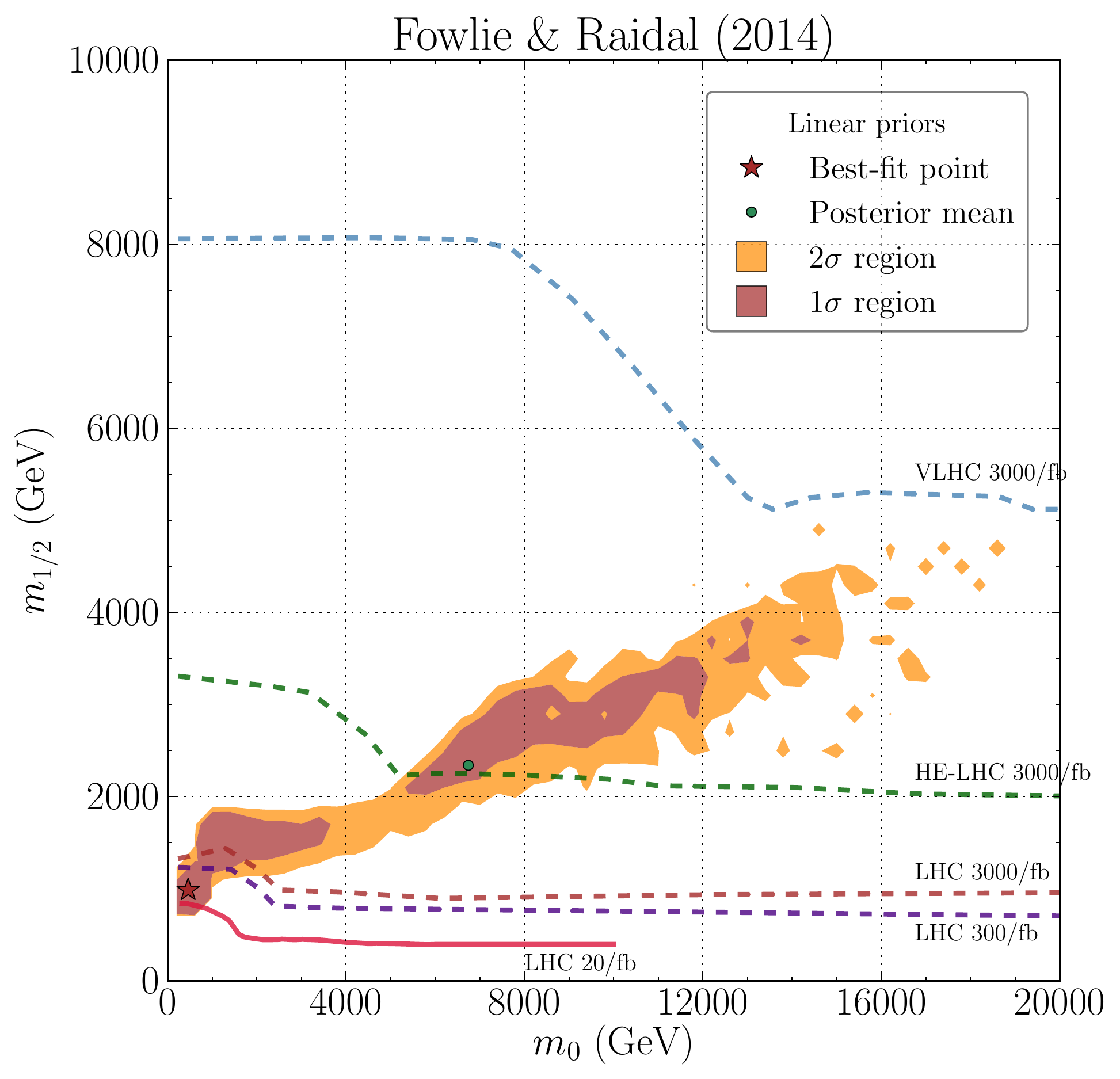}%
}%
\subfloat[H][Credible regions on the \pat plane with linear priors. ]{\label{fig:a0tanb_linear}%
\includegraphics[height=0.49\linewidth]{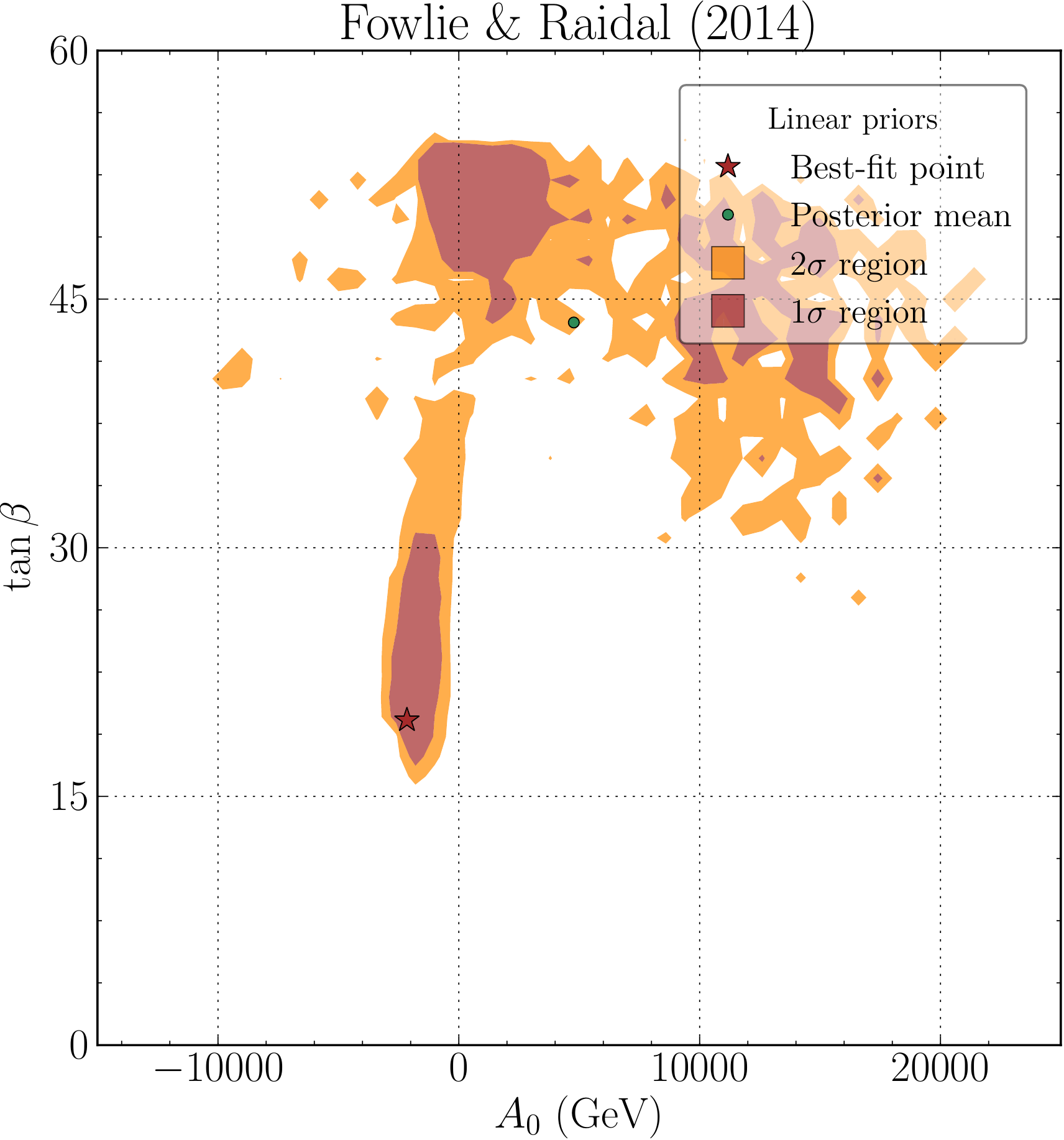}%
}%
\caption{The \pmm  and \pat planes of the CMSSM showing the $68\%$ (red) and $95\%$ (orange) credible regions of the marginalized posterior. The expected discovery potentials of future hadron colliders are also shown.}
\label{fig:linear}
\end{figure}

We plot the credible regions of the marginalized posterior on the \pmm plane in \reffig{fig:m0m12:i}. One can identify by eye three modes on the \pmm plane, which are characterized by the mechanism by which dark matter is annihilated in 
the early Universe.  The three dominant mechanisms present are:
\begin{itemize}
 \item \Stauc with a bino-like neutralino at $1\sigma$ at light soft-breaking masses; $\mzero\sim1\tev$ and $\mhalf\sim1\tev$. Because  $\mneut{1}\simeq\ms{\tau}$, \stauc is significant, and proceeds via stau-tau-bino gauge interaction vertices in $s$- and $t$-channel
 diagrams. Such diagrams are suppressed if neutralinos are heavy, by a heavy $t$-channel stau or an off-shell  $s$-channel tau.  
 
 \item $A$-funnel $s$-channel annihilation at $2\sigma$ at intermediate  soft-breaking masses; $0.5\tev\lesssim\mzero\lesssim2\tev$ and $\mhalf\sim1.5\tev$. Because $\mneut{1} \simeq \ma/2$, neutralinos annihilate via an $s$-channel pseudo-scalar Higgs (which avoids helicity
  suppression, because the vertex is a Yukawa, and $p$-wave suppression, because it couples via the $L=0$ partial wave).  Annihilation is via a Higgs-higgsino-gaugino vertex;
  the neutralino must have a mixed composition, though is dominantly bino-like. Satisfying $\mneut{1} \simeq \ma/2$ becomes impossible with heavy soft-breaking masses in the CMSSM, 
  though is possible in MSSM models, in which \ma is a free parameter\cite{Fowlie:2013oua}. Note that with lighter \ma, supersymmetric contributions to \bsmm  are appreciable.
  
  \item The so-called ``$1\tev$ higgsino region''\cite{Roszkowski:2009sm,Fowlie:2013oua,Kowalska:2013hha} at $2\sigma$ at heavy  soft-breaking masses; $5\tev\lesssim\mzero\lesssim9\tev$ and $2\tev\lesssim\mhalf\lesssim3\tev$. The higgsino-like neutralino annihilates via a $t$-channel charged higgsino to $WW$ with higgsino-charged higgsino-gauge boson vertices or coannihilates with a higgsino-like chargino to an $ff^\prime$ pair. This region is similar in this regard to the focus-point. With heavier $\mu\gtrsim1\tev$, the relic density is increased. Because $\mu\gg M_W$, the mass splitting between the higgsino-like chargino
  and higgsino-like neutralinos is negligible. The higgsino-like chargino is ``parasitic'' in the language of \refcite{Profumo:2013yn} and increases the relic density.  With lighter $\mu\lesssim1\tev$,  the Higgs mass is lighter than its measured value.  
  
  Because the neutralino is higgsino-like, the spin-independent scattering cross-section with a proton, \sigsip, is enhanced via an $s$-channel Higgs diagram, with higgsino-bino-Higgs vertices.  Although with $\mu<0$, light and heavy Higgs bosons  could destructively interfere, the heavy Higgs is so heavy that significant cancellations rarely occur. This region is disfavored by LUX, which at $\mneut{1}\simeq1\tev$ restricts $\sigsip \lesssim 10^{-44}\,\text{cm}^2$.
  
\end{itemize}
These mechanisms are not disjoint;  $A$-funnel annihilation is present, though somewhat off-resonance, at heavier \mhalf above the ``$1\tev$ higgsino region.''  A  line of fine-tuned solutions exist from the $A$-funnel mode along the top of the ``$1\tev$ higgsino region.'' Although $\mneut{1}-\ma/2$ is increasing along this line, so too is the pseudoscalar's width. Regardless of the dark matter annihilation mechanism and regardless of our priors, we find that the neutralino mass must be less than $1160\gev$ at $95\%$ (one tail).  In \reffig{fig:m0m12_linear}, we see that the sizes of the $A$-funnel  and``$1\tev$ higgsino region''  on the \pmm plane are considerably increased if one uses linear priors, rather than logarithmic priors, and both are present at $1\sigma$ rather than  $2\sigma$.

Other mechanisms, \eg $Z$/$h$ resonances and bulk annihilation, which could annihilate dark matter at a sufficient rate, are excluded by LHC direct searches for sparticles.  Our posterior is, by eye, in good agreement with that in \refcite{Kowalska:2013hha}, which omits direct detection experiments, with $\Delta\chi^2$ plots in \refcite{Buchmueller:2013rsa} and with crude results in \refcite{Kadastik:2011aa}. \refcite{Kowalska:2013hha} identifies a ``focus-point'' region at $\mzero\sim4\tev$ and  $\mhalf\sim1\tev$. Whilst we see a few such points in \reffig{fig:m0m12:ii} at the beginning of the ``$1\tev$ higgsino region,'' their combined posterior weight is negligible, and they are absent in \reffig{fig:m0m12:i}.  The ``focus-point'' struggles to produce $\mh\sim125\gev$ and is disfavored by the LUX experiment. It might have an appreciable posterior weight in \refcite{Kowalska:2013hha} because that analysis included a $3\gev$ theory error (rather than our $2\gev$ theory error) in the \mh calculation and omitted direct detection experiments. 

\refcite{Kadastik:2011aa} identified stop-coannihilation as a 
possible dark matter annihilation mechanism. Because the \texttt{MultiNest-2.18} algorithm found few such solutions (although we confirm that such points with reasonable $\chi^2$ exist), the stop-coannihilation regions are fine-tuned so much so that their posterior weight is negligible. We plot stop-coannihilation solutions in \reffig{fig:m0m12:ii}. 

\begin{figure}[t]
\centering
\subfloat[][Credible regions.]{\label{fig:a0tanb:i}
\includegraphics[width=0.49\linewidth]{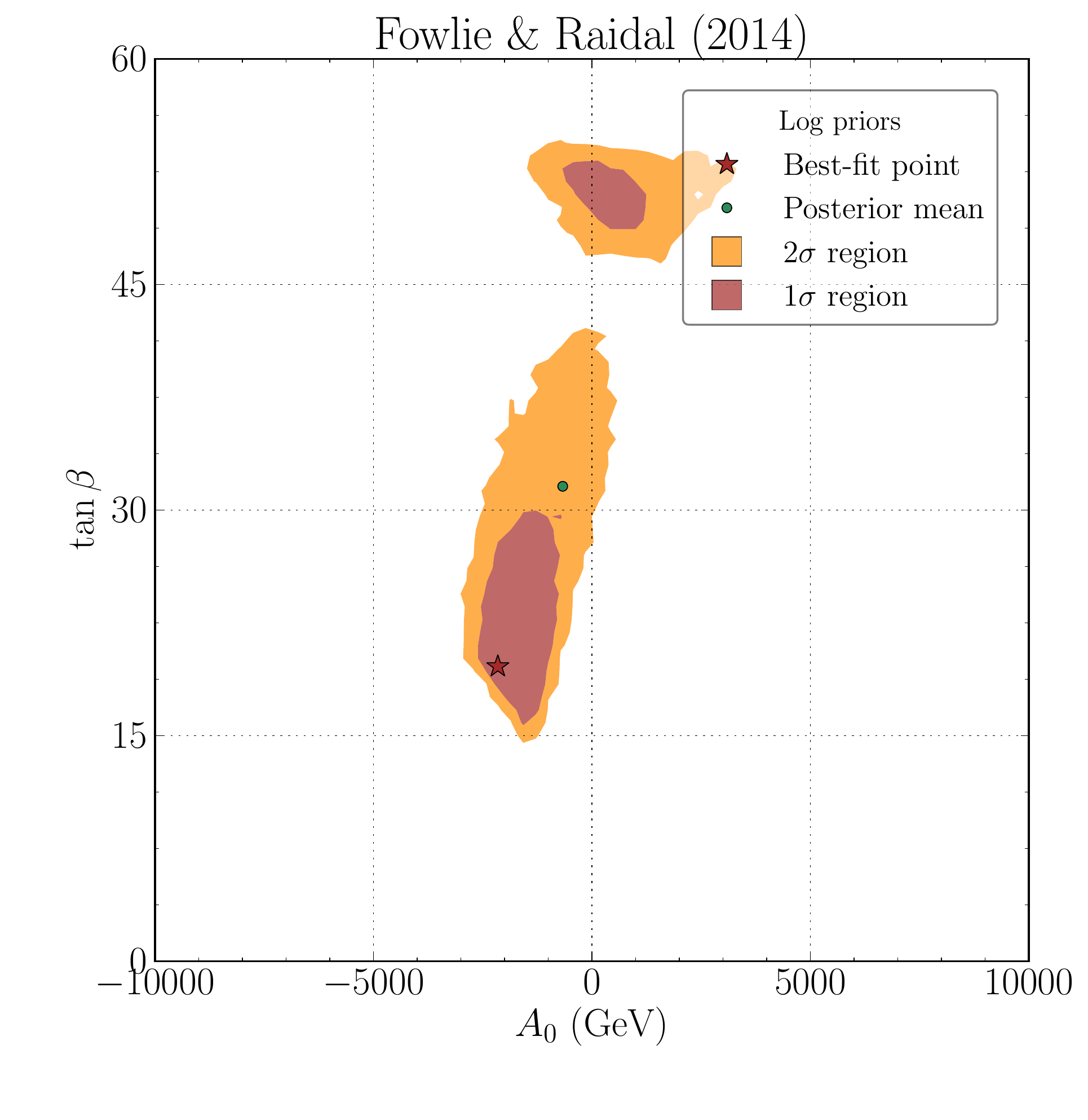}
}
\subfloat[][Scatter plot illustrating dark matter annihilation mechanisms. ]{\label{fig:a0tanb:ii}
\includegraphics[width=0.49\linewidth]{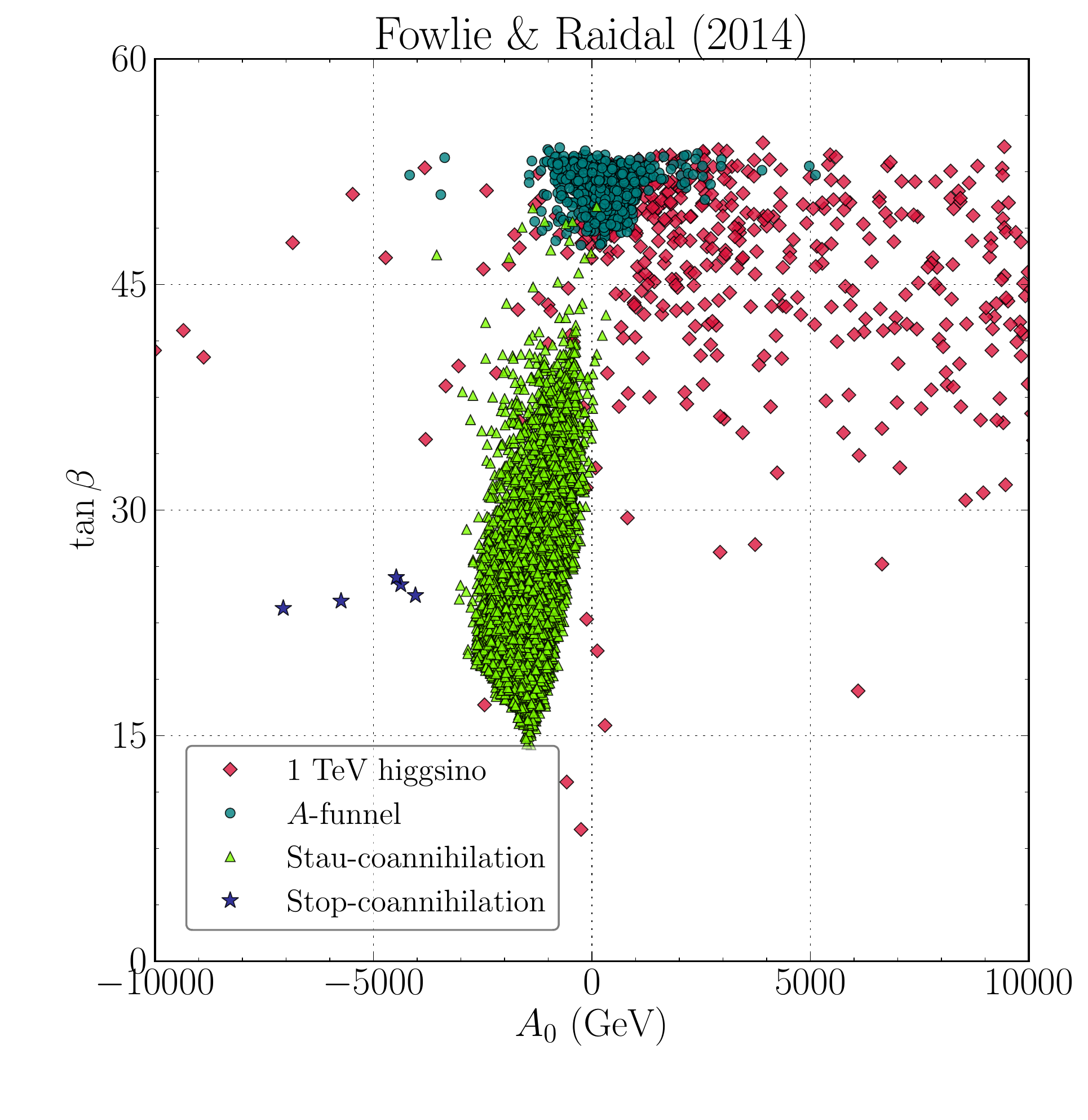}
}
\caption{The \pmm plane of the CMSSM showing: \protect\subref{fig:a0tanb:i} The $68\%$ (red) and $95\%$ (orange) credible regions of the marginalized posterior and \protect\subref{fig:a0tanb:ii} A scatter of the points with appreciable posterior weight colored by their dominant dark matter annihilation mechanism.}
\label{fig:a0tanb}
\end{figure}

On the \pat plane in \reffig{fig:a0tanb},  we identify two modes that correspond to the \stauc and $A$-funnel regions on the \pmm plane. The \stauc region prefers $\tanb\lesssim30$ and negative \azero to achieve the required mass degeneracy and stop mixing. Within the \stauc region, $\mu>0$ is preferred to enhance \damu with light sparticle loops. The $A$-funnel requires $\tanb\simeq50$ and prefers $\azero\gtrsim1.5\tev$ so that the pseudo-scalar Higgs is sufficiently light.  

Our posterior on this plane is, by eye, in agreement with $\Delta\chi^2$ plots in \refcite{Buchmueller:2013rsa}, but in poor agreement with that in \refcite{Kowalska:2013hha}. In each case, one can identify a \stauc region at small \tanb. We, however, fail to see modes at  $\azero\sim\pm8\tev$ present in \refcite{Kowalska:2013hha}, corresponding to the ``$1\tev$ higgsino region.'' The higgsino-like neutralino is disfavored by the LUX direct search for dark matter. No direct detection constraint was applied in \refcite{Kowalska:2013hha}, though its potential impact was investigated, and found to be substantial in the ``$1\tev$ higgsino region,'' in agreement with our findings. That the ``$1\tev$ higgsino region'' is present on the \pmm plane but not on the \pat plane is a result of marginalization. Acceptable $\mu\simeq1\tev$ solutions result from a moderate range of \pat, but a rather restricted range of \pmm. On the \pmm plane in the ``$1\tev$ higgsino region,'' the posterior weight is enhanced by the many \pat solutions, whereas, on the \pat plane in the ``$1\tev$ higgsino region,'' the posterior weight is the sum of few \pmm solutions. 

\begin{figure}[t]
\centering
\subfloat[][Credible regions with logarithmic priors.]{\label{fig:dd_log}
\includegraphics[width=0.49\linewidth]{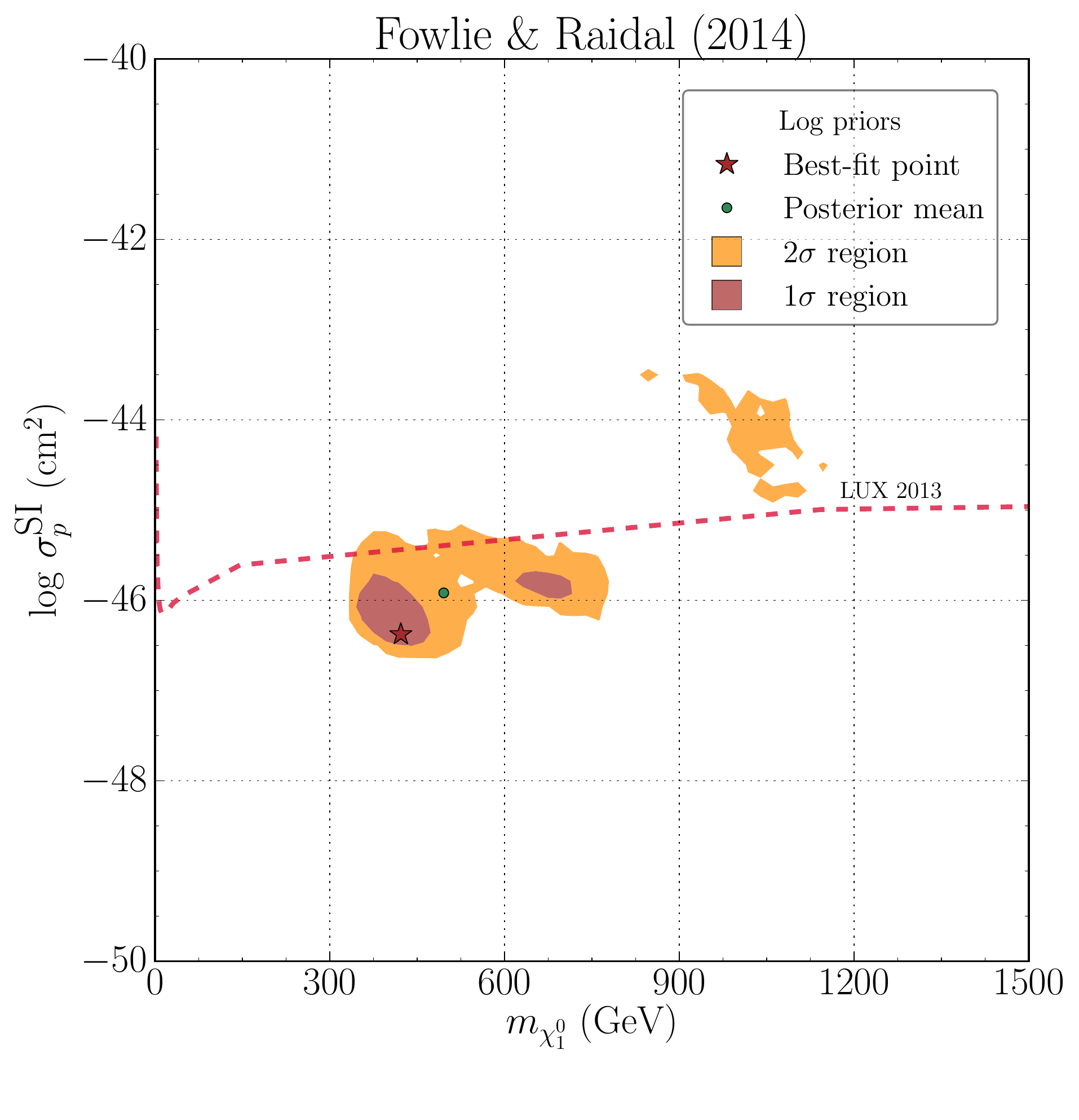}
}
\subfloat[][Credible regions with linear priors. ]{\label{fig:dd_linear}
\includegraphics[width=0.49\linewidth]{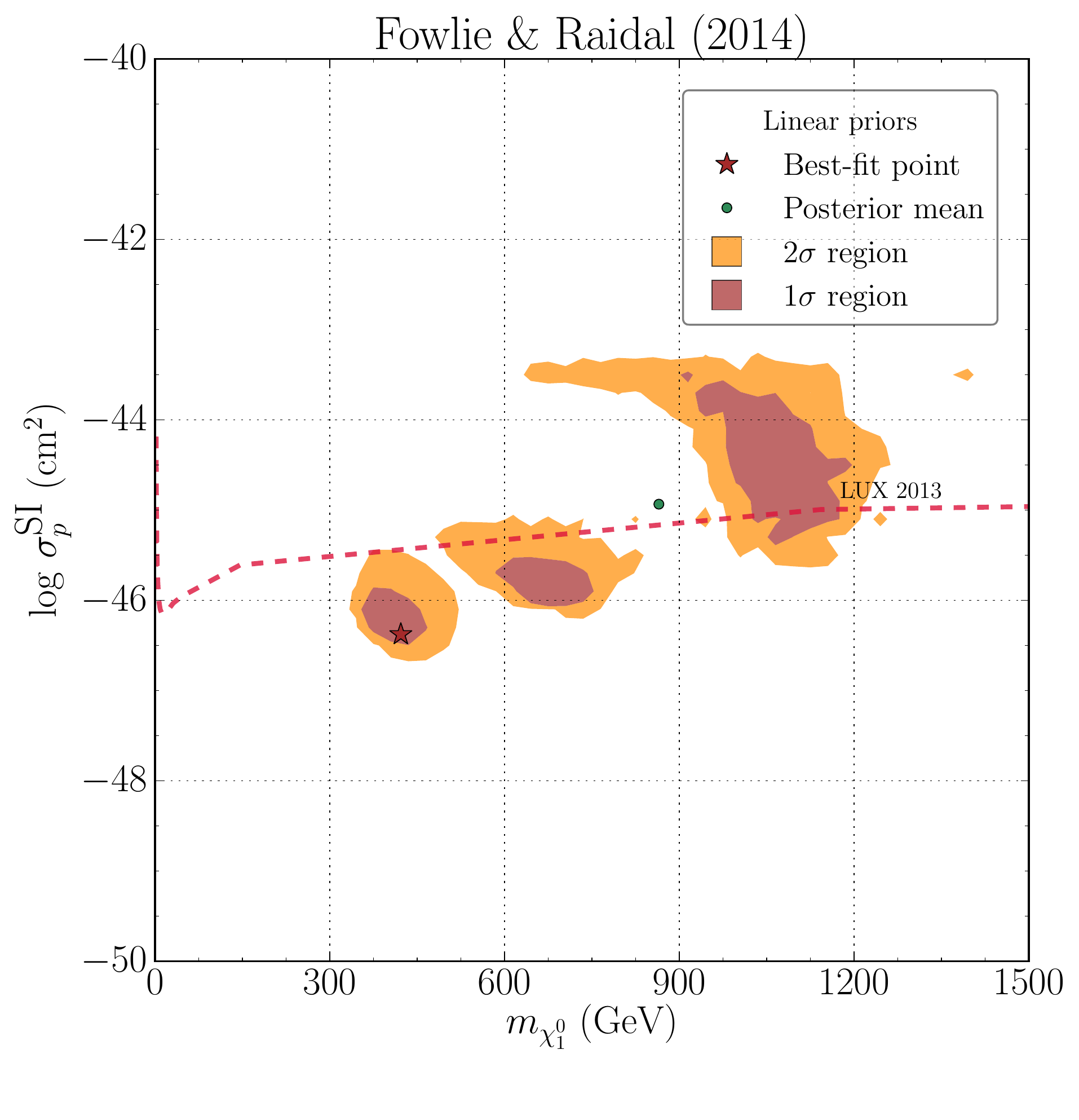}
}
\caption{The $68\%$ (red) and $95\%$ (orange) credible regions of the marginalized posterior on the \pcs plane. The $90\%$ exclusion contour from LUX is also shown.}
\label{fig:dd}
\end{figure}

The ``$1\tev$ higgsino region'' is, however, visible on the \pat plane with linear priors  in \reffig{fig:a0tanb_linear} with $\tanb\sim45$ and $10\tev\lesssim\azero\lesssim20\tev$. Large \azero and \tanb are preferred to achieve $\mu\sim1\tev$ and to insure that radiative electroweak symmetry breaking with $m^2_{H_u}<0$ is achieved, in spite of such heavy \mzero.  
Because the stop-coannihilation mechanism requires large $a_t$ at the electroweak scale to split the stops so that the lightest is degenerate with the neutralino, it prefers large negative \azero.
 
On the \pcs plane in \reffig{fig:dd}, which is relevant to direct searches for dark matter, we again identify three modes characterized by their dark matter annihilation mechanism: the \stauc region at $\mneut{1}\sim500\gev$,  the $A$-funnel at $\mneut{1}\sim700\gev$ and the ``$1\tev$ higgsino region.'' As anticipated, the scattering cross-section  is largest in the ``$1\tev$ higgsino region,'' because the higgsino-bino-Higgs vertices are enhanced by the neutralino's composition. Linear priors (\reffig{fig:dd_linear}) favor the ``$1\tev$ higgsino region.''  The LUX limit was, of course, included in our likelihood function; \textit{prima facie} it is surprising that the ``$1\tev$ higgsino region'' lies above that limit. This is because there is significant theoretical uncertainty in the scattering cross-sections. There is an appreciable chance that regions with \textit{calculated} scattering cross-sections that are above the LUX limit in fact have \textit{true} scattering cross-sections that are below the LUX limit. Our Bayesian methodology and results reflect this fact. 

 To conclude, let us summarize the impact of the experimental measurements in our likelihood function in \reftable{tab:data}:
\begin{itemize}
 \item If the relic density, \abund, is to be sufficiently small, the neutralino's mass or composition must be fine-tuned to enhance a particular annihilation mechanism. This selects narrow regions of parameter space.
 \item The Higgs mass, $\mh\sim125\gev$, prefers intermediate soft-breaking masses or $\lesssim1\tev$ soft-breaking masses and maximal stop-mixing. Higgsino-like neutralinos with either $\mzero\lesssim5\tev$ or $\mzero\gtrsim10\tev$ are disfavored. The $W$-boson mass, \mw,  further disfavors $\mh\gtrsim125\gev$, even if sparticle-loop contributions are negligible\cite{Heinemeyer:2013dia}.
 \item The null result of the LUX direct search for dark matter disfavors higgsino-like neutralinos.
 \item The effect of LHC direct searches  is somewhat trivial; a low-mass portion of the \pmm plane is excluded.
 \item With heavy sparticles,  sparticle loop contributions to electroweak precision observables and $b$-physics are insubstantial, and their impact is somewhat limited. A positive sign of the Higgs parameter is favored by \damu in the \stauc region, in which sparticles are light. Furthermore, \bsmm disfavors light pseudo-scalar Higgs masses, which could enhance \bsmm, which in turn disfavors $A$-funnel annihilation if the neutralino is light.
\end{itemize}

\section{\label{sec:accessible}Results and discussion }
For concreteness, we say that a CMSSM point is discoverable if, were nature described by that CMSSM point,  it is expected that the SM background hypothesis could be rejected at $5\sigma$.  We assume that two simplified channels\cite{Alves:2011wf} studied in \refcite{Cohen:2013xda} describe the hadron collider phenomenology of the CMSSM's favored regions:
\begin{itemize}
 \item If gluinos and squarks are light, we assume a simplified gluino-squark-neutralino channel in which gluinos and squarks are pair produced and decay to a neutralino and a quark or quark pair. Because the neutralinos must be less than approximately $1\tev$ at $95\%$ from our posterior, one can expect a discovery potential of $\ms{g}\simeq\ms{q}\lesssim2.7\tev$( $\ms{g}\simeq\ms{q}\lesssim3.0\tev$) at the LHC with  $300\invfb$ ($3000\invfb$), $\ms{g}\simeq\ms{q}\lesssim6.6\tev$ at the HE-LHC and $\ms{g}\simeq\ms{q}\lesssim15\tev$ at the VLHC with $3000\invfb$\cite{Cohen:2013xda}.
  \item If squarks are heavy, we assume a simplified gluino-neutralino channel in which gluinos are pair produced and decay to a neutralino and a quark pair via an off-shell squark. Because the neutralinos must be less than approximately $1\tev$ at $95\%$ from our posterior, one can expect a discovery potential of $\ms{g}\lesssim1.9\tev$ ($\ms{g}\lesssim2.2\tev$) at the LHC with  $300\invfb$ ($3000\invfb$), $\ms{g}\lesssim5.0\tev$ at the HE-LHC and $\ms{g}\lesssim11\tev$ at the VLHC with $3000\invfb$\cite{Cohen:2013xda}. 
\end{itemize}
We neglect electroweak production, which is unlikely to significantly extend the discovery potential at a hadron collider.

The discovery potential of the various experiments is shown on the \pmm planes in \reffig{fig:m0m12} and \reffig{fig:linear} for logarithmic and linear priors, respectively. 
For logarithmic priors, the LHC with $300\invfb$ could discover all of the CMSSM's \stauc region and a light fraction of the $A$-funnel region 
 while the HE-LHC could discover all of the $A$-funnel. The ``$1\tev$ higgsino region,''  however, is entirely beyond the reach of the LHC, 
 and partially beyond that of the HE-LHC. For linear priors, large part of the $A$-funnel region remains uncovered also by the HE-LHC. 
However, for the VLHC our conclusion is the same for both choices of the prior: the CMSSM can be discovered for all favored dark matter annihilation mechanisms. 
We stress once more the complementarity between collider searches and direct detection, as seen in \reffig{fig:dd}.
Improving direct detection constraints and/or reducing the associated nuclear uncertainties could entirely exclude the ``$1\tev$ higgsino region.'' 

It is evident that our results, which we calculate from \refeq{eq:pfinal}, are somewhat sensitive to our choice of priors.  We calculated our probabilities with linear priors and logarithmic priors for the soft-breaking masses. Linear priors weight linear intervals evenly, \ie $\priorf{x}\propto\text{constant.}$ Whilst different investigators might have a spectrum of prior beliefs for the CMSSM's parameters, all investigators ought to make identical conclusions from their posteriors, if the experimental data is ``strong enough.'' Although we believe that, because they are scale invariant, logarithmic priors for the soft-breaking masses are the best choice, linear priors are not unreasonable. We checked that a $\pm10\%$ systematic uncertainty in the expected hadron collider discovery reach had limited impact compared to that of our prior choice. In the  paragraphs that follow, we quote the range of probabilities obtained from logarithmic and linear priors rounded to the nearest whole $5\%$. Note that null results from future direct detection and \bsmm experiments would disfavor the $A$-funnel and ``$1\tev$ higgsino region.'' Because those regions are disfavored by logarithmic priors, if in the future such experiments were to obtain null results, the probabilities calculated with logarithmic priors would be more accurate than those with linear priors.

\begin{table}[t]
\begin{center}
\begin{ruledtabular}
\begin{tabular}{lp{2.5cm}p{2.5cm}p{2.5cm}p{2.5cm}}
Experiment & \multicolumn{2}{c}{\shortstack{Probability of discovering SUSY,\\ given data, CMSSM correct model }}  & \multicolumn{2}{c}{\shortstack{\ldots and  given that previous experiment did not\\ discover the CMSSM}}\\
\hline
& Log priors & Linear priors & Log priors & Linear priors\\
\hline

LHC $300\invfb$ & $73\%$  & $15\%$ & --- & --- \\

LHC $3000\invfb$& $76\%$ & $17\%$  & $9\%$  &  $2\%$\\

HE-LHC $3000\invfb$& $96\%$  & $45\%$ & $83\%$ &  $34\%$ \\

VLHC $3000\invfb$ & $100\%$ & $100\%$ & $100\%$& $100\%$\\

\end{tabular}
\end{ruledtabular}
\end{center}
\caption{\label{tab:prob} 
Probability of discovering SUSY at various experiments with logarithmic and linear priors. All probabilities are conditioned on experimental data thus far obtained and on the assumption that the CMSSM is the correct model.  In the second set of two columns, the probability is also conditioned on the proposition that the previous collider experiments in the table failed to discover SUSY.}
\end{table}

We find that from \refeq{eq:pfinal}, the probability that  the CMSSM could be discovered at the VLHC  is always $100\%$, at the HE-LHC, $45\dash95\%$ and at the LHC with $300\invfb$ ($3000\invfb$), $15\dash75\%$ ($15\dash75\%$). All probabilities are conditional on the experimental data obtained thus far and on the assumption that the CMSSM is the true model. 

It is insightful, however, to consider the probability of a discovery at a hypothetical experiment given that all experiments \textit{including hypothetical experiments that would have been performed by that time} did not make a discovery, \eg
\begin{align}
&\pg{\text{Discoverable at HE-LHC}}{\data,\model,\text{No LHC experiments make a discovery}} \\\nonumber
&=\frac{\pg{\text{Discoverable at HE-LHC},\text{Not discoverable at LHC}}{\data,\model}}{\pg{\text{Not discoverable at LHC}}{\data,\model}}\\\nonumber
&=\frac{\pg{\text{Discoverable at HE-LHC}}{\data,\model} - \pg{\text{Discoverable at LHC}}{\data,\model}}{1-\pg{\text{Discoverable at LHC}}{\data,\model}}.
\end{align}
With this caveat the probability of discovering the CMSSM at the LHC with $3000\invfb$ is  $1\dash10\%$ and at the HE-LHC, $35\dash85\%$. Our probabilities suggest that the HE-LHC is rather likely to discover SUSY, assuming that the CMSSM is the correct model. The VLHC should have the final word on the CMSSM; we find with near certainty that if the CMSSM is the correct model, the VLHC will discover SUSY. The LHC $3000\invfb$ is unlikely to discover the CMSSM, if it was not already discovered with $300\invfb$. We summarize all probabilities in \reftable{tab:prob}.

 \begin{figure}[t]
\centering
\subfloat[][The \pmm plane.]{\label{fig:model par}
\includegraphics[width=0.49\linewidth]{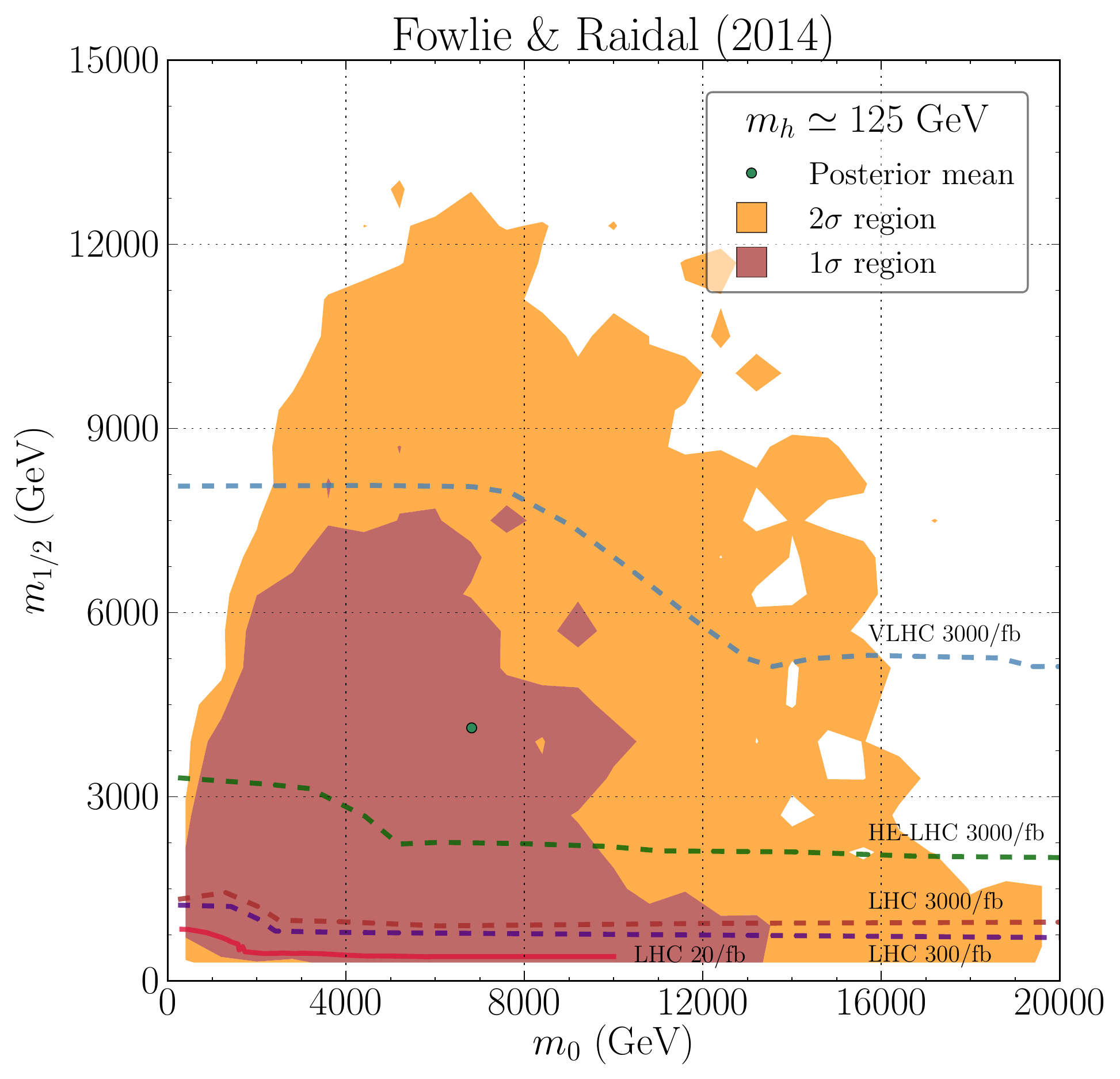}
}
\subfloat[][The stop-gluino mass plane. ]{\label{fig:physmass}
\includegraphics[width=0.49\linewidth]{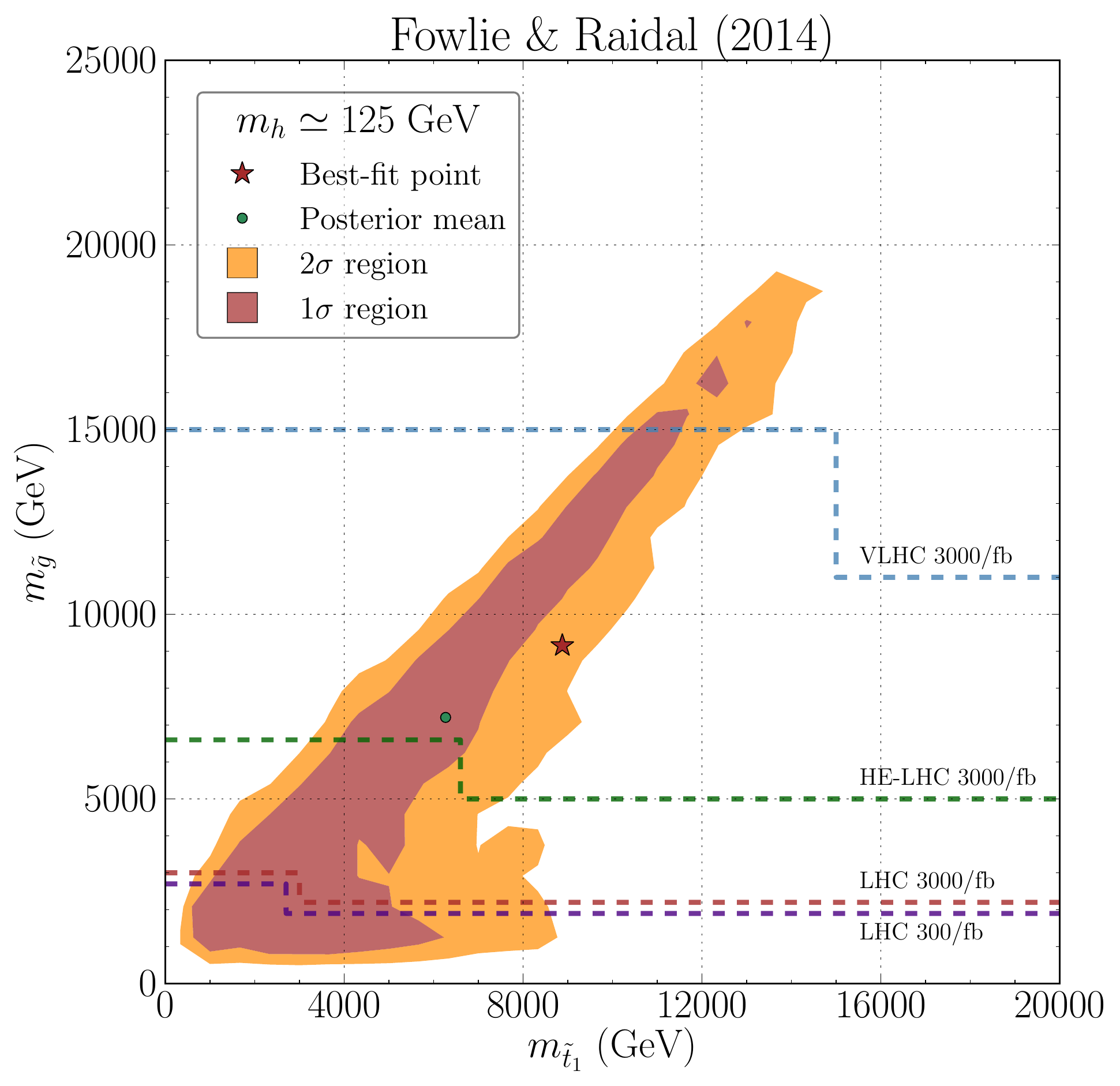}
}
\caption{The  CMSSM  $68\%$ (red) and $95\%$ (orange) credible regions of the marginalized posterior with only the measurement of the Higgs mass in our likelihood function and with logarithmic priors. The expected discovery potentials of future hadron colliders are also shown.}
\label{fig:Higgs}
\end{figure}

Our probabilities were, of course, dependent on our assumption that the dark matter is entirely neutralino. We recalculated the probabilities with  the minimal assumption that the CMSSM accounted for only the Higgs boson mass, by including only the measurement of the Higgs mass in our likelihood function. We permitted, \textit{a priori,} $\tanb\simeq1$, which is disfavored by \eg the dark matter relic density measurement, but reduces the tree-level Higgs mass. In all our results, however, we vetoed points for which Yukawa couplings were non-perturbative below the GUT scale or points that were otherwise unphysical. 

The resulting posterior is plotted in \reffig{fig:Higgs}  for the \pmm  and for stop-gluino mass planes.\footnote{A similar result was first found in \refcite{Kowalska:2013hha}.} The effect of giving up the requirement of obtaining the observed amount of DM is evident: part of the high mass parameter region remains uncovered by all the planned colliders.
Whilst the credible regions suggest that the VLHC has an appreciable ($85\%$) probability of discovering the CMSSM, this conclusion is somewhat dependent on our priors.  Soft-breaking masses of greater than $\sim10\tev$ typically result in a Higgs mass that is greater than its measured value, but are also disfavored by our logarithmic priors. With linear priors for \mzero and \mhalf, that probability is moderate ($50\%$).  The results plotted in \reffig{fig:Higgs} demonstrate that DM annihilation processes do constrain SUSY 
parameters significantly.

We assumed that $\mhalf\le10\tev$ and $\mzero\le20\tev$ in our priors in \reftable{tab:priors}; this choice omitted no credible parts of parameter space or DM annihilation mechanisms. To check whether our probabilities were sensitive to this assumption, we reduced our prior ranges to $\mhalf\le5\tev$ and $\mzero\le10\tev$ and recalculated our probabilities. With logarithmic priors, the probabilities changed by $\sim1$ percentage point, because the $95\%$ credible region on the \pmm plane in \reffig{fig:m0m12} was within $\mhalf\le5\tev$ and $\mzero\le10\tev$. With linear priors, the $95\%$ credible region on the \pmm plane in \reffig{fig:linear} exceeded $\mzero\le10\tev$ in the ``$1\tev$ higgsino region;'' applying the prior range $\mzero\le10\tev$ truncated the ``$1\tev$ higgsino region.'' The probabilities for the LHC increased by $\sim5$ percentage points, whereas that for the HE-LHC decreased by $\sim20$ percentage points. The probability for the VLHC was unchanged. This indicates sensitivity to the prior range for \mzero. We believe, however, that our prior range for \mzero was a fair choice, because, unlike the reduced prior range, our choice omitted no viable DM annihilation mechanisms. 

\section{Conclusions}
Whilst in the near future experiments at the LHC are expected to continue at \roots{14}, discussions are underway for a high-energy proton-proton collider.  Two preliminary ideas are the \roots{100} VLHC and the \roots{33} HE-LHC.  We calculated the Bayesian posterior density of the CMSSM's parameter space, given experimental data from the LHC, LUX and Planck. Our result was in agreement with similar previous analyses and could be understood with reference to possible dark matter annihilation mechanisms. With Bayesian statistics, we calculated the probabilities that the LHC, HE-LHC and VLHC discover SUSY in the future, assuming that nature is described by the CMSSM and given the experimental data from the LHC, LUX and Planck (\eg assuming that dark matter is entirely neutralino). 

We found that the LHC with $300\invfb$ at \roots{14} has a $15\dash75\%$ probability of discovering SUSY. If that experiment does not discover SUSY, the probability of discovering SUSY with  $3000\invfb$ is merely $1\dash10\%$. Were SUSY to remain undetected at the LHC, the HE-LHC would have a $35\dash85\%$ probability of discovering SUSY with $3000\invfb$. The VLHC, on the other hand,  ought to be definitive;  it has a $100\%$ probability of discovering SUSY with $3000\invfb$. All probabilities, summarized in \reftable{tab:prob},
 are conditional on the experimental data obtained thus far and on the assumption that the CMSSM is the true model. We remind the reader that in Bayesian statistics, probability is a numerical measure of belief, and is sensitive to one's prior beliefs, \eg our priors for the CMSSM in \reftable{tab:priors}, though we believe our priors were fair. 
Nevertheless, our finding that the VLHC has a $100\%$ probability of discovering SUSY with $3000\invfb$ was independent of our choice of prior for the soft-breaking masses. 
 We checked that our conclusions were robust with respect to systematic errors in the expected performance of the collider experiments. 
 
 Stated qualitatively, our conclusions are that there is a fair probability of discovering the CMSSM at the LHC with $300\invfb$ at \roots{14}. If that search is unsuccessful, the CMSSM is unlikely to be discovered in $3000\invfb$. A \roots{33} HE-LHC with $3000\invfb$ would be likely to discover the CMSSM, and, should it be unsuccessful, a \roots{100} VLHC would definitively discover the CMSSM.
 
 Our results were primarily determined by the CMSSM's predictions for the Higgs mass and the DM abundance, which we assumed to be entirely neutralinos.  If, however, DM has a different origin, the CMSSM could be unreachable even at the VLHC. Our conclusions are applicable to only the CMSSM; in relaxed models, one can tune the pseudo-scalar resonance so that heavy neutralinos are annihilated at a sufficient rate.

\begin{acknowledgements}
We thank M.~Kadastik for helpful discussions. 
This work was supported in part by grants IUT23-6, CERN+,  and by the European Union
through the European Regional Development Fund and by ERDF project 3.2.0304.11-0313
Estonian Scientific Computing Infrastructure (ETAIS).
\end{acknowledgements}

\bibliography{main}

\begin{thebibliography}{65}
\expandafter\ifx\csname natexlab\endcsname\relax\def\natexlab#1{#1}\fi
\expandafter\ifx\csname bibnamefont\endcsname\relax
  \def\bibnamefont#1{#1}\fi
\expandafter\ifx\csname bibfnamefont\endcsname\relax
  \def\bibfnamefont#1{#1}\fi
\expandafter\ifx\csname citenamefont\endcsname\relax
  \def\citenamefont#1{#1}\fi
\expandafter\ifx\csname url\endcsname\relax
  \def\url#1{\texttt{#1}}\fi
\expandafter\ifx\csname urlprefix\endcsname\relax\def\urlprefix{URL }\fi
\providecommand{\bibinfo}[2]{#2}
\providecommand{\eprint}[2][]{\url{#2}}

\bibitem[{\citenamefont{Salam and Strathdee}(1974)}]{Salam:1974yz}
\bibinfo{author}{\bibfnamefont{A.}~\bibnamefont{Salam}} \bibnamefont{and}
  \bibinfo{author}{\bibfnamefont{J.}~\bibnamefont{Strathdee}},
  \bibinfo{journal}{Nucl.Phys.} \textbf{\bibinfo{volume}{B76}},
  \bibinfo{pages}{477} (\bibinfo{year}{1974}).

\bibitem[{\citenamefont{Haber and Kane}(1985)}]{Haber:1984rc}
\bibinfo{author}{\bibfnamefont{H.~E.} \bibnamefont{Haber}} \bibnamefont{and}
  \bibinfo{author}{\bibfnamefont{G.~L.} \bibnamefont{Kane}},
  \bibinfo{journal}{Phys.Rept.} \textbf{\bibinfo{volume}{117}},
  \bibinfo{pages}{75} (\bibinfo{year}{1985}).

\bibitem[{\citenamefont{Nilles}(1984)}]{Nilles:1983ge}
\bibinfo{author}{\bibfnamefont{H.~P.} \bibnamefont{Nilles}},
  \bibinfo{journal}{Phys.Rept.} \textbf{\bibinfo{volume}{110}},
  \bibinfo{pages}{1} (\bibinfo{year}{1984}).

\bibitem[{\citenamefont{Georgi and Glashow}(1974)}]{Georgi:1974sy}
\bibinfo{author}{\bibfnamefont{H.}~\bibnamefont{Georgi}} \bibnamefont{and}
  \bibinfo{author}{\bibfnamefont{S.}~\bibnamefont{Glashow}},
  \bibinfo{journal}{Phys.Rev.Lett.} \textbf{\bibinfo{volume}{32}},
  \bibinfo{pages}{438} (\bibinfo{year}{1974}).

\bibitem[{\citenamefont{Dimopoulos and Georgi}(1981)}]{Dimopoulos:1981zb}
\bibinfo{author}{\bibfnamefont{S.}~\bibnamefont{Dimopoulos}} \bibnamefont{and}
  \bibinfo{author}{\bibfnamefont{H.}~\bibnamefont{Georgi}},
  \bibinfo{journal}{Nucl.Phys.} \textbf{\bibinfo{volume}{B193}},
  \bibinfo{pages}{150} (\bibinfo{year}{1981}).

\bibitem[{\citenamefont{Jungman et~al.}(1996)\citenamefont{Jungman,
  Kamionkowski, and Griest}}]{Jungman:1995df}
\bibinfo{author}{\bibfnamefont{G.}~\bibnamefont{Jungman}},
  \bibinfo{author}{\bibfnamefont{M.}~\bibnamefont{Kamionkowski}},
  \bibnamefont{and} \bibinfo{author}{\bibfnamefont{K.}~\bibnamefont{Griest}},
  \bibinfo{journal}{Phys.Rept.} \textbf{\bibinfo{volume}{267}},
  \bibinfo{pages}{195} (\bibinfo{year}{1996}), \eprint{hep-ph/9506380}.

\bibitem[{\citenamefont{Assmann et~al.}(2010)\citenamefont{Assmann, Bailey,
  Brüning, Dominguez~Sanchez, de~Rijk, Jimenez, Myers, Rossi, Tavian, Todesco
  et~al.}}]{Assmann:1284326}
\bibinfo{author}{\bibfnamefont{R.}~\bibnamefont{Assmann}},
  \bibinfo{author}{\bibfnamefont{R.}~\bibnamefont{Bailey}},
  \bibinfo{author}{\bibfnamefont{O.}~\bibnamefont{Brüning}},
  \bibinfo{author}{\bibfnamefont{O.}~\bibnamefont{Dominguez~Sanchez}},
  \bibinfo{author}{\bibfnamefont{G.}~\bibnamefont{de~Rijk}},
  \bibinfo{author}{\bibfnamefont{J.~M.} \bibnamefont{Jimenez}},
  \bibinfo{author}{\bibfnamefont{S.}~\bibnamefont{Myers}},
  \bibinfo{author}{\bibfnamefont{L.}~\bibnamefont{Rossi}},
  \bibinfo{author}{\bibfnamefont{L.}~\bibnamefont{Tavian}},
  \bibinfo{author}{\bibfnamefont{E.}~\bibnamefont{Todesco}},
  \bibnamefont{et~al.}, \bibinfo{type}{Tech. Rep.}
  \bibinfo{number}{CERN-ATS-2010-177}, \bibinfo{institution}{CERN},
  \bibinfo{address}{Geneva} (\bibinfo{year}{2010}).

\bibitem[{\citenamefont{{Future Circular Collider Kickoff
  Meeting}}(2014)}]{FCC}
\bibinfo{author}{\bibnamefont{{Future Circular Collider Kickoff Meeting}}}
  (\bibinfo{address}{University of Geneva, Geneva, Switzerland},
  \bibinfo{year}{2014}),
  \urlprefix\url{http://indico.cern.ch/conferenceDisplay.py?confId=282344}.

\bibitem[{\citenamefont{LeCompte and Martin}(2011)}]{LeCompte:2011cn}
\bibinfo{author}{\bibfnamefont{T.~J.} \bibnamefont{LeCompte}} \bibnamefont{and}
  \bibinfo{author}{\bibfnamefont{S.~P.} \bibnamefont{Martin}},
  \bibinfo{journal}{Phys.Rev.} \textbf{\bibinfo{volume}{D84}},
  \bibinfo{pages}{015004} (\bibinfo{year}{2011}), \eprint{1105.4304}.

\bibitem[{\citenamefont{Fayet}(1975)}]{Fayet:1974pd}
\bibinfo{author}{\bibfnamefont{P.}~\bibnamefont{Fayet}},
  \bibinfo{journal}{Nucl.Phys.} \textbf{\bibinfo{volume}{B90}},
  \bibinfo{pages}{104} (\bibinfo{year}{1975}).

\bibitem[{\citenamefont{Ellwanger et~al.}(2010)\citenamefont{Ellwanger,
  Hugonie, and Teixeira}}]{Ellwanger:2009dp}
\bibinfo{author}{\bibfnamefont{U.}~\bibnamefont{Ellwanger}},
  \bibinfo{author}{\bibfnamefont{C.}~\bibnamefont{Hugonie}}, \bibnamefont{and}
  \bibinfo{author}{\bibfnamefont{A.~M.} \bibnamefont{Teixeira}},
  \bibinfo{journal}{Phys.Rept.} \textbf{\bibinfo{volume}{496}},
  \bibinfo{pages}{1} (\bibinfo{year}{2010}), \eprint{0910.1785}.

\bibitem[{\citenamefont{Trotta}(2008)}]{Trotta:2008qt}
\bibinfo{author}{\bibfnamefont{R.}~\bibnamefont{Trotta}},
  \bibinfo{journal}{Contemp.Phys.} \textbf{\bibinfo{volume}{49}},
  \bibinfo{pages}{71} (\bibinfo{year}{2008}), \eprint{0803.4089}.

\bibitem[{\citenamefont{Chamseddine et~al.}(1982)\citenamefont{Chamseddine,
  Arnowitt, and Nath}}]{Chamseddine:1982jx}
\bibinfo{author}{\bibfnamefont{A.~H.} \bibnamefont{Chamseddine}},
  \bibinfo{author}{\bibfnamefont{R.~L.} \bibnamefont{Arnowitt}},
  \bibnamefont{and} \bibinfo{author}{\bibfnamefont{P.}~\bibnamefont{Nath}},
  \bibinfo{journal}{Phys.Rev.Lett.} \textbf{\bibinfo{volume}{49}},
  \bibinfo{pages}{970} (\bibinfo{year}{1982}).

\bibitem[{\citenamefont{Arnowitt and Nath}(1992)}]{Arnowitt:1992aq}
\bibinfo{author}{\bibfnamefont{R.~L.} \bibnamefont{Arnowitt}} \bibnamefont{and}
  \bibinfo{author}{\bibfnamefont{P.}~\bibnamefont{Nath}},
  \bibinfo{journal}{Phys.Rev.Lett.} \textbf{\bibinfo{volume}{69}},
  \bibinfo{pages}{725} (\bibinfo{year}{1992}).

\bibitem[{\citenamefont{Kane et~al.}(1994)\citenamefont{Kane, Kolda,
  Roszkowski, and Wells}}]{Kane:1993td}
\bibinfo{author}{\bibfnamefont{G.~L.} \bibnamefont{Kane}},
  \bibinfo{author}{\bibfnamefont{C.~F.} \bibnamefont{Kolda}},
  \bibinfo{author}{\bibfnamefont{L.}~\bibnamefont{Roszkowski}},
  \bibnamefont{and} \bibinfo{author}{\bibfnamefont{J.~D.} \bibnamefont{Wells}},
  \bibinfo{journal}{Phys.Rev.} \textbf{\bibinfo{volume}{D49}},
  \bibinfo{pages}{6173} (\bibinfo{year}{1994}), \eprint{hep-ph/9312272}.

\bibitem[{\citenamefont{Beskidt et~al.}(2014)\citenamefont{Beskidt, de~Boer,
  and Kazakov}}]{Beskidt:2014oea}
\bibinfo{author}{\bibfnamefont{C.}~\bibnamefont{Beskidt}},
  \bibinfo{author}{\bibfnamefont{W.}~\bibnamefont{de~Boer}}, \bibnamefont{and}
  \bibinfo{author}{\bibfnamefont{D.}~\bibnamefont{Kazakov}}
  (\bibinfo{year}{2014}), \eprint{1402.4650}.

\bibitem[{\citenamefont{Cohen and Wacker}(2013)}]{Cohen:2013kna}
\bibinfo{author}{\bibfnamefont{T.}~\bibnamefont{Cohen}} \bibnamefont{and}
  \bibinfo{author}{\bibfnamefont{J.~G.} \bibnamefont{Wacker}},
  \bibinfo{journal}{JHEP} \textbf{\bibinfo{volume}{1309}}, \bibinfo{pages}{061}
  (\bibinfo{year}{2013}), \eprint{1305.2914}.

\bibitem[{\citenamefont{Henrot-Versillé
  et~al.}(2013)\citenamefont{Henrot-Versillé, Lafaye, Plehn, Rauch, Zerwas
  et~al.}}]{Henrot-Versille:2013yma}
\bibinfo{author}{\bibfnamefont{S.}~\bibnamefont{Henrot-Versillé}},
  \bibinfo{author}{\bibfnamefont{R.}~\bibnamefont{Lafaye}},
  \bibinfo{author}{\bibfnamefont{T.}~\bibnamefont{Plehn}},
  \bibinfo{author}{\bibfnamefont{M.}~\bibnamefont{Rauch}},
  \bibinfo{author}{\bibfnamefont{D.}~\bibnamefont{Zerwas}},
  \bibnamefont{et~al.} (\bibinfo{year}{2013}), \eprint{1309.6958}.

\bibitem[{\citenamefont{Buchmueller et~al.}(2013)\citenamefont{Buchmueller,
  Cavanaugh, De~Roeck, Dolan, Ellis et~al.}}]{Buchmueller:2013rsa}
\bibinfo{author}{\bibfnamefont{O.}~\bibnamefont{Buchmueller}},
  \bibinfo{author}{\bibfnamefont{R.}~\bibnamefont{Cavanaugh}},
  \bibinfo{author}{\bibfnamefont{A.}~\bibnamefont{De~Roeck}},
  \bibinfo{author}{\bibfnamefont{M.}~\bibnamefont{Dolan}},
  \bibinfo{author}{\bibfnamefont{J.}~\bibnamefont{Ellis}}, \bibnamefont{et~al.}
  (\bibinfo{year}{2013}), \eprint{1312.5250}.

\bibitem[{\citenamefont{Bechtle et~al.}(2013)\citenamefont{Bechtle, Desch,
  Dreiner, Hamer, Krämer et~al.}}]{Bechtle:2013mda}
\bibinfo{author}{\bibfnamefont{P.}~\bibnamefont{Bechtle}},
  \bibinfo{author}{\bibfnamefont{K.}~\bibnamefont{Desch}},
  \bibinfo{author}{\bibfnamefont{H.~K.} \bibnamefont{Dreiner}},
  \bibinfo{author}{\bibfnamefont{M.}~\bibnamefont{Hamer}},
  \bibinfo{author}{\bibfnamefont{M.}~\bibnamefont{Krämer}},
  \bibnamefont{et~al.} (\bibinfo{year}{2013}), \eprint{1310.3045}.

\bibitem[{\citenamefont{Kowalska et~al.}(2013)\citenamefont{Kowalska,
  Roszkowski, and Sessolo}}]{Kowalska:2013hha}
\bibinfo{author}{\bibfnamefont{K.}~\bibnamefont{Kowalska}},
  \bibinfo{author}{\bibfnamefont{L.}~\bibnamefont{Roszkowski}},
  \bibnamefont{and} \bibinfo{author}{\bibfnamefont{E.~M.}
  \bibnamefont{Sessolo}}, \bibinfo{journal}{JHEP}
  \textbf{\bibinfo{volume}{1306}}, \bibinfo{pages}{078} (\bibinfo{year}{2013}),
  \eprint{1302.5956}.

\bibitem[{\citenamefont{Fowlie et~al.}(2012{\natexlab{a}})\citenamefont{Fowlie,
  Kazana, Kowalska, Munir, Roszkowski et~al.}}]{Fowlie:2012im}
\bibinfo{author}{\bibfnamefont{A.}~\bibnamefont{Fowlie}},
  \bibinfo{author}{\bibfnamefont{M.}~\bibnamefont{Kazana}},
  \bibinfo{author}{\bibfnamefont{K.}~\bibnamefont{Kowalska}},
  \bibinfo{author}{\bibfnamefont{S.}~\bibnamefont{Munir}},
  \bibinfo{author}{\bibfnamefont{L.}~\bibnamefont{Roszkowski}},
  \bibnamefont{et~al.}, \bibinfo{journal}{Phys.Rev.}
  \textbf{\bibinfo{volume}{D86}}, \bibinfo{pages}{075010}
  (\bibinfo{year}{2012}{\natexlab{a}}), \eprint{1206.0264}.

\bibitem[{\citenamefont{Roszkowski et~al.}(2012)\citenamefont{Roszkowski,
  Sessolo, and Tsai}}]{Roszkowski:2012uf}
\bibinfo{author}{\bibfnamefont{L.}~\bibnamefont{Roszkowski}},
  \bibinfo{author}{\bibfnamefont{E.~M.} \bibnamefont{Sessolo}},
  \bibnamefont{and} \bibinfo{author}{\bibfnamefont{Y.-L.~S.}
  \bibnamefont{Tsai}}, \bibinfo{journal}{Phys.Rev.}
  \textbf{\bibinfo{volume}{D86}}, \bibinfo{pages}{095005}
  (\bibinfo{year}{2012}), \eprint{1202.1503}.

\bibitem[{\citenamefont{Strege et~al.}(2012)\citenamefont{Strege, Bertone,
  Feroz, Fornasa, de~Austri et~al.}}]{Strege:2012bt}
\bibinfo{author}{\bibfnamefont{C.}~\bibnamefont{Strege}},
  \bibinfo{author}{\bibfnamefont{G.}~\bibnamefont{Bertone}},
  \bibinfo{author}{\bibfnamefont{F.}~\bibnamefont{Feroz}},
  \bibinfo{author}{\bibfnamefont{M.}~\bibnamefont{Fornasa}},
  \bibinfo{author}{\bibfnamefont{R.~R.} \bibnamefont{de~Austri}},
  \bibnamefont{et~al.} (\bibinfo{year}{2012}), \eprint{1212.2636}.

\bibitem[{\citenamefont{Allanach et~al.}(2011)\citenamefont{Allanach, Khoo,
  Lester, and Williams}}]{Allanach:2011wi}
\bibinfo{author}{\bibfnamefont{B.}~\bibnamefont{Allanach}},
  \bibinfo{author}{\bibfnamefont{T.}~\bibnamefont{Khoo}},
  \bibinfo{author}{\bibfnamefont{C.}~\bibnamefont{Lester}}, \bibnamefont{and}
  \bibinfo{author}{\bibfnamefont{S.}~\bibnamefont{Williams}},
  \bibinfo{journal}{JHEP} \textbf{\bibinfo{volume}{1106}}, \bibinfo{pages}{035}
  (\bibinfo{year}{2011}), \eprint{1103.0969}.

\bibitem[{\citenamefont{Akula et~al.}(2012)\citenamefont{Akula, Nath, and
  Peim}}]{Akula:2012kk}
\bibinfo{author}{\bibfnamefont{S.}~\bibnamefont{Akula}},
  \bibinfo{author}{\bibfnamefont{P.}~\bibnamefont{Nath}}, \bibnamefont{and}
  \bibinfo{author}{\bibfnamefont{G.}~\bibnamefont{Peim}},
  \bibinfo{journal}{Phys.Lett.} \textbf{\bibinfo{volume}{B717}},
  \bibinfo{pages}{188} (\bibinfo{year}{2012}), \eprint{1207.1839}.

\bibitem[{\citenamefont{Cabrera et~al.}(2012)\citenamefont{Cabrera, Casas, and
  de~Austri}}]{Cabrera:2012vu}
\bibinfo{author}{\bibfnamefont{M.~E.} \bibnamefont{Cabrera}},
  \bibinfo{author}{\bibfnamefont{J.~A.} \bibnamefont{Casas}}, \bibnamefont{and}
  \bibinfo{author}{\bibfnamefont{R.~R.} \bibnamefont{de~Austri}}
  (\bibinfo{year}{2012}), \eprint{1212.4821}.

\bibitem[{\citenamefont{Feroz et~al.}(2009)\citenamefont{Feroz, Hobson, and
  Bridges}}]{Feroz:2008xx}
\bibinfo{author}{\bibfnamefont{F.}~\bibnamefont{Feroz}},
  \bibinfo{author}{\bibfnamefont{M.}~\bibnamefont{Hobson}}, \bibnamefont{and}
  \bibinfo{author}{\bibfnamefont{M.}~\bibnamefont{Bridges}},
  \bibinfo{journal}{Mon.Not.Roy.Astron.Soc.} \textbf{\bibinfo{volume}{398}},
  \bibinfo{pages}{1601} (\bibinfo{year}{2009}), \eprint{0809.3437}.

\bibitem[{\citenamefont{Buchner et~al.}(2014)\citenamefont{Buchner,
  Georgakakis, Nandra, Hsu, Rangel et~al.}}]{Buchner:2014nha}
\bibinfo{author}{\bibfnamefont{J.}~\bibnamefont{Buchner}},
  \bibinfo{author}{\bibfnamefont{A.}~\bibnamefont{Georgakakis}},
  \bibinfo{author}{\bibfnamefont{K.}~\bibnamefont{Nandra}},
  \bibinfo{author}{\bibfnamefont{L.}~\bibnamefont{Hsu}},
  \bibinfo{author}{\bibfnamefont{C.}~\bibnamefont{Rangel}},
  \bibnamefont{et~al.} (\bibinfo{year}{2014}), \eprint{1402.0004}.

\bibitem[{\citenamefont{Fowlie et~al.}(2012{\natexlab{b}})\citenamefont{Fowlie,
  Kalinowski, Kazana, Roszkowski, and Tsai}}]{Fowlie:2011mb}
\bibinfo{author}{\bibfnamefont{A.}~\bibnamefont{Fowlie}},
  \bibinfo{author}{\bibfnamefont{A.}~\bibnamefont{Kalinowski}},
  \bibinfo{author}{\bibfnamefont{M.}~\bibnamefont{Kazana}},
  \bibinfo{author}{\bibfnamefont{L.}~\bibnamefont{Roszkowski}},
  \bibnamefont{and} \bibinfo{author}{\bibfnamefont{Y.-L.~S.}
  \bibnamefont{Tsai}}, \bibinfo{journal}{Phys.Rev.}
  \textbf{\bibinfo{volume}{D85}}, \bibinfo{pages}{075012}
  (\bibinfo{year}{2012}{\natexlab{b}}), \eprint{1111.6098}.

\bibitem[{\citenamefont{Belanger et~al.}(2011)\citenamefont{Belanger, Boudjema,
  Brun, Pukhov, Rosier-Lees et~al.}}]{Belanger:2010gh}
\bibinfo{author}{\bibfnamefont{G.}~\bibnamefont{Belanger}},
  \bibinfo{author}{\bibfnamefont{F.}~\bibnamefont{Boudjema}},
  \bibinfo{author}{\bibfnamefont{P.}~\bibnamefont{Brun}},
  \bibinfo{author}{\bibfnamefont{A.}~\bibnamefont{Pukhov}},
  \bibinfo{author}{\bibfnamefont{S.}~\bibnamefont{Rosier-Lees}},
  \bibnamefont{et~al.}, \bibinfo{journal}{Comput.Phys.Commun.}
  \textbf{\bibinfo{volume}{182}}, \bibinfo{pages}{842} (\bibinfo{year}{2011}),
  \eprint{1004.1092}.

\bibitem[{\citenamefont{Belanger et~al.}(2007)\citenamefont{Belanger, Boudjema,
  Pukhov, and Semenov}}]{Belanger:2006is}
\bibinfo{author}{\bibfnamefont{G.}~\bibnamefont{Belanger}},
  \bibinfo{author}{\bibfnamefont{F.}~\bibnamefont{Boudjema}},
  \bibinfo{author}{\bibfnamefont{A.}~\bibnamefont{Pukhov}}, \bibnamefont{and}
  \bibinfo{author}{\bibfnamefont{A.}~\bibnamefont{Semenov}},
  \bibinfo{journal}{Comput.Phys.Commun.} \textbf{\bibinfo{volume}{176}},
  \bibinfo{pages}{367} (\bibinfo{year}{2007}), \eprint{hep-ph/0607059}.

\bibitem[{\citenamefont{Heinemeyer et~al.}(2000)\citenamefont{Heinemeyer,
  Hollik, and Weiglein}}]{Heinemeyer:1998yj}
\bibinfo{author}{\bibfnamefont{S.}~\bibnamefont{Heinemeyer}},
  \bibinfo{author}{\bibfnamefont{W.}~\bibnamefont{Hollik}}, \bibnamefont{and}
  \bibinfo{author}{\bibfnamefont{G.}~\bibnamefont{Weiglein}},
  \bibinfo{journal}{Comput.Phys.Commun.} \textbf{\bibinfo{volume}{124}},
  \bibinfo{pages}{76} (\bibinfo{year}{2000}), \eprint{hep-ph/9812320}.

\bibitem[{\citenamefont{Heinemeyer et~al.}(1999)\citenamefont{Heinemeyer,
  Hollik, and Weiglein}}]{Heinemeyer:1998np}
\bibinfo{author}{\bibfnamefont{S.}~\bibnamefont{Heinemeyer}},
  \bibinfo{author}{\bibfnamefont{W.}~\bibnamefont{Hollik}}, \bibnamefont{and}
  \bibinfo{author}{\bibfnamefont{G.}~\bibnamefont{Weiglein}},
  \bibinfo{journal}{Eur.Phys.J.} \textbf{\bibinfo{volume}{C9}},
  \bibinfo{pages}{343} (\bibinfo{year}{1999}), \eprint{hep-ph/9812472}.

\bibitem[{\citenamefont{Degrassi et~al.}(2003)\citenamefont{Degrassi,
  Heinemeyer, Hollik, Slavich, and Weiglein}}]{Degrassi:2002fi}
\bibinfo{author}{\bibfnamefont{G.}~\bibnamefont{Degrassi}},
  \bibinfo{author}{\bibfnamefont{S.}~\bibnamefont{Heinemeyer}},
  \bibinfo{author}{\bibfnamefont{W.}~\bibnamefont{Hollik}},
  \bibinfo{author}{\bibfnamefont{P.}~\bibnamefont{Slavich}}, \bibnamefont{and}
  \bibinfo{author}{\bibfnamefont{G.}~\bibnamefont{Weiglein}},
  \bibinfo{journal}{Eur.Phys.J.} \textbf{\bibinfo{volume}{C28}},
  \bibinfo{pages}{133} (\bibinfo{year}{2003}), \eprint{hep-ph/0212020}.

\bibitem[{\citenamefont{Frank et~al.}(2007)\citenamefont{Frank, Hahn,
  Heinemeyer, Hollik, Rzehak et~al.}}]{Frank:2006yh}
\bibinfo{author}{\bibfnamefont{M.}~\bibnamefont{Frank}},
  \bibinfo{author}{\bibfnamefont{T.}~\bibnamefont{Hahn}},
  \bibinfo{author}{\bibfnamefont{S.}~\bibnamefont{Heinemeyer}},
  \bibinfo{author}{\bibfnamefont{W.}~\bibnamefont{Hollik}},
  \bibinfo{author}{\bibfnamefont{H.}~\bibnamefont{Rzehak}},
  \bibnamefont{et~al.}, \bibinfo{journal}{JHEP}
  \textbf{\bibinfo{volume}{0702}}, \bibinfo{pages}{047} (\bibinfo{year}{2007}),
  \eprint{hep-ph/0611326}.

\bibitem[{\citenamefont{Allanach}(2002)}]{Allanach:2001kg}
\bibinfo{author}{\bibfnamefont{B.}~\bibnamefont{Allanach}},
  \bibinfo{journal}{Comput.Phys.Commun.} \textbf{\bibinfo{volume}{143}},
  \bibinfo{pages}{305} (\bibinfo{year}{2002}), \eprint{hep-ph/0104145}.

\bibitem[{ATL(2013)}]{ATLAS-CONF-2013-047}
\bibinfo{type}{Tech. Rep.} \bibinfo{number}{ATLAS-CONF-2013-047},
  \bibinfo{institution}{CERN}, \bibinfo{address}{Geneva}
  (\bibinfo{year}{2013}).

\bibitem[{\citenamefont{Allanach}(2011)}]{Allanach:2011ut}
\bibinfo{author}{\bibfnamefont{B.}~\bibnamefont{Allanach}},
  \bibinfo{journal}{Phys.Rev.} \textbf{\bibinfo{volume}{D83}},
  \bibinfo{pages}{095019} (\bibinfo{year}{2011}), \eprint{1102.3149}.

\bibitem[{\citenamefont{Bechtle et~al.}(2011)\citenamefont{Bechtle, Sarrazin,
  Desch, Dreiner, Wienemann et~al.}}]{Bechtle:2011dm}
\bibinfo{author}{\bibfnamefont{P.}~\bibnamefont{Bechtle}},
  \bibinfo{author}{\bibfnamefont{B.}~\bibnamefont{Sarrazin}},
  \bibinfo{author}{\bibfnamefont{K.}~\bibnamefont{Desch}},
  \bibinfo{author}{\bibfnamefont{H.~K.} \bibnamefont{Dreiner}},
  \bibinfo{author}{\bibfnamefont{P.}~\bibnamefont{Wienemann}},
  \bibnamefont{et~al.}, \bibinfo{journal}{Phys.Rev.}
  \textbf{\bibinfo{volume}{D84}}, \bibinfo{pages}{011701}
  (\bibinfo{year}{2011}), \eprint{1102.4693}.

\bibitem[{\citenamefont{Akerib et~al.}(2013)}]{Akerib:2013tjd}
\bibinfo{author}{\bibfnamefont{D.}~\bibnamefont{Akerib}} \bibnamefont{et~al.}
  (\bibinfo{collaboration}{LUX Collaboration}) (\bibinfo{year}{2013}),
  \eprint{1310.8214}.

\bibitem[{\citenamefont{Ellis et~al.}(2008)\citenamefont{Ellis, Olive, and
  Savage}}]{Ellis:2008hf}
\bibinfo{author}{\bibfnamefont{J.~R.} \bibnamefont{Ellis}},
  \bibinfo{author}{\bibfnamefont{K.~A.} \bibnamefont{Olive}}, \bibnamefont{and}
  \bibinfo{author}{\bibfnamefont{C.}~\bibnamefont{Savage}},
  \bibinfo{journal}{Phys.Rev.} \textbf{\bibinfo{volume}{D77}},
  \bibinfo{pages}{065026} (\bibinfo{year}{2008}), \eprint{0801.3656}.

\bibitem[{\citenamefont{Ade et~al.}(2013)}]{Ade:2013zuv}
\bibinfo{author}{\bibfnamefont{P.}~\bibnamefont{Ade}} \bibnamefont{et~al.}
  (\bibinfo{collaboration}{Planck Collaboration}) (\bibinfo{year}{2013}),
  \eprint{1303.5076}.

\bibitem[{\citenamefont{Allanach
  et~al.}(2004{\natexlab{a}})\citenamefont{Allanach, Belanger, Boudjema, and
  Pukhov}}]{Allanach:2004jy}
\bibinfo{author}{\bibfnamefont{B.~C.} \bibnamefont{Allanach}},
  \bibinfo{author}{\bibfnamefont{G.}~\bibnamefont{Belanger}},
  \bibinfo{author}{\bibfnamefont{F.}~\bibnamefont{Boudjema}}, \bibnamefont{and}
  \bibinfo{author}{\bibfnamefont{A.}~\bibnamefont{Pukhov}}, pp.
  \bibinfo{pages}{961--964} (\bibinfo{year}{2004}{\natexlab{a}}),
  \eprint{hep-ph/0410049}.

\bibitem[{\citenamefont{Allanach
  et~al.}(2004{\natexlab{b}})\citenamefont{Allanach, Belanger, Boudjema,
  Pukhov, and Porod}}]{Allanach:2004jh}
\bibinfo{author}{\bibfnamefont{B.}~\bibnamefont{Allanach}},
  \bibinfo{author}{\bibfnamefont{G.}~\bibnamefont{Belanger}},
  \bibinfo{author}{\bibfnamefont{F.}~\bibnamefont{Boudjema}},
  \bibinfo{author}{\bibfnamefont{A.}~\bibnamefont{Pukhov}}, \bibnamefont{and}
  \bibinfo{author}{\bibfnamefont{W.}~\bibnamefont{Porod}}
  (\bibinfo{year}{2004}{\natexlab{b}}), \eprint{hep-ph/0402161}.

\bibitem[{\citenamefont{Beringer et~al.}(2012)}]{Beringer:1900zz}
\bibinfo{author}{\bibfnamefont{J.}~\bibnamefont{Beringer}} \bibnamefont{et~al.}
  (\bibinfo{collaboration}{Particle Data Group}), \bibinfo{journal}{Phys.Rev.}
  \textbf{\bibinfo{volume}{D86}}, \bibinfo{pages}{010001}
  (\bibinfo{year}{2012}).

\bibitem[{\citenamefont{Chatrchyan et~al.}(2012)}]{Chatrchyan:2012ufa}
\bibinfo{author}{\bibfnamefont{S.}~\bibnamefont{Chatrchyan}}
  \bibnamefont{et~al.} (\bibinfo{collaboration}{CMS Collaboration}),
  \bibinfo{journal}{Phys.Lett.} \textbf{\bibinfo{volume}{B716}},
  \bibinfo{pages}{30} (\bibinfo{year}{2012}), \eprint{1207.7235}.

\bibitem[{\citenamefont{Aad et~al.}(2012)}]{Aad:2012tfa}
\bibinfo{author}{\bibfnamefont{G.}~\bibnamefont{Aad}} \bibnamefont{et~al.}
  (\bibinfo{collaboration}{ATLAS Collaboration}), \bibinfo{journal}{Phys.Lett.}
  \textbf{\bibinfo{volume}{B716}}, \bibinfo{pages}{1} (\bibinfo{year}{2012}),
  \eprint{1207.7214}.

\bibitem[{\citenamefont{Allanach
  et~al.}(2004{\natexlab{c}})\citenamefont{Allanach, Djouadi, Kneur, Porod, and
  Slavich}}]{Allanach:2004rh}
\bibinfo{author}{\bibfnamefont{B.}~\bibnamefont{Allanach}},
  \bibinfo{author}{\bibfnamefont{A.}~\bibnamefont{Djouadi}},
  \bibinfo{author}{\bibfnamefont{J.}~\bibnamefont{Kneur}},
  \bibinfo{author}{\bibfnamefont{W.}~\bibnamefont{Porod}}, \bibnamefont{and}
  \bibinfo{author}{\bibfnamefont{P.}~\bibnamefont{Slavich}},
  \bibinfo{journal}{JHEP} \textbf{\bibinfo{volume}{0409}}, \bibinfo{pages}{044}
  (\bibinfo{year}{2004}{\natexlab{c}}), \eprint{hep-ph/0406166}.

\bibitem[{\citenamefont{Heinemeyer et~al.}(2006)\citenamefont{Heinemeyer,
  Hollik, and Weiglein}}]{Heinemeyer:2004gx}
\bibinfo{author}{\bibfnamefont{S.}~\bibnamefont{Heinemeyer}},
  \bibinfo{author}{\bibfnamefont{W.}~\bibnamefont{Hollik}}, \bibnamefont{and}
  \bibinfo{author}{\bibfnamefont{G.}~\bibnamefont{Weiglein}},
  \bibinfo{journal}{Phys.Rept.} \textbf{\bibinfo{volume}{425}},
  \bibinfo{pages}{265} (\bibinfo{year}{2006}), \eprint{hep-ph/0412214}.

\bibitem[{\citenamefont{Trotta et~al.}(2008)\citenamefont{Trotta, Feroz,
  Hobson, Roszkowski, and Ruiz~de Austri}}]{Trotta:2008bp}
\bibinfo{author}{\bibfnamefont{R.}~\bibnamefont{Trotta}},
  \bibinfo{author}{\bibfnamefont{F.}~\bibnamefont{Feroz}},
  \bibinfo{author}{\bibfnamefont{M.~P.} \bibnamefont{Hobson}},
  \bibinfo{author}{\bibfnamefont{L.}~\bibnamefont{Roszkowski}},
  \bibnamefont{and} \bibinfo{author}{\bibfnamefont{R.}~\bibnamefont{Ruiz~de
  Austri}}, \bibinfo{journal}{JHEP} \textbf{\bibinfo{volume}{0812}},
  \bibinfo{pages}{024} (\bibinfo{year}{2008}), \eprint{0809.3792}.

\bibitem[{\citenamefont{de~Austri et~al.}(2006)\citenamefont{de~Austri, Trotta,
  and Roszkowski}}]{deAustri:2006pe}
\bibinfo{author}{\bibfnamefont{R.~R.} \bibnamefont{de~Austri}},
  \bibinfo{author}{\bibfnamefont{R.}~\bibnamefont{Trotta}}, \bibnamefont{and}
  \bibinfo{author}{\bibfnamefont{L.}~\bibnamefont{Roszkowski}},
  \bibinfo{journal}{JHEP} \textbf{\bibinfo{volume}{0605}}, \bibinfo{pages}{002}
  (\bibinfo{year}{2006}), \eprint{hep-ph/0602028}.

\bibitem[{\citenamefont{Amhis et~al.}(2012)}]{Amhis:2012bh}
\bibinfo{author}{\bibfnamefont{Y.}~\bibnamefont{Amhis}} \bibnamefont{et~al.}
  (\bibinfo{collaboration}{Heavy Flavor Averaging Group})
  (\bibinfo{year}{2012}), \eprint{1207.1158}.

\bibitem[{\citenamefont{Misiak et~al.}(2007)\citenamefont{Misiak, Asatrian,
  Bieri, Czakon, Czarnecki et~al.}}]{Misiak:2006zs}
\bibinfo{author}{\bibfnamefont{M.}~\bibnamefont{Misiak}},
  \bibinfo{author}{\bibfnamefont{H.}~\bibnamefont{Asatrian}},
  \bibinfo{author}{\bibfnamefont{K.}~\bibnamefont{Bieri}},
  \bibinfo{author}{\bibfnamefont{M.}~\bibnamefont{Czakon}},
  \bibinfo{author}{\bibfnamefont{A.}~\bibnamefont{Czarnecki}},
  \bibnamefont{et~al.}, \bibinfo{journal}{Phys.Rev.Lett.}
  \textbf{\bibinfo{volume}{98}}, \bibinfo{pages}{022002}
  (\bibinfo{year}{2007}), \eprint{hep-ph/0609232}.

\bibitem[{\citenamefont{Buchmueller et~al.}(2012)\citenamefont{Buchmueller,
  Cavanaugh, De~Roeck, Dolan, Ellis et~al.}}]{Buchmueller:2011sw}
\bibinfo{author}{\bibfnamefont{O.}~\bibnamefont{Buchmueller}},
  \bibinfo{author}{\bibfnamefont{R.}~\bibnamefont{Cavanaugh}},
  \bibinfo{author}{\bibfnamefont{A.}~\bibnamefont{De~Roeck}},
  \bibinfo{author}{\bibfnamefont{M.}~\bibnamefont{Dolan}},
  \bibinfo{author}{\bibfnamefont{J.}~\bibnamefont{Ellis}},
  \bibnamefont{et~al.}, \bibinfo{journal}{Eur.Phys.J.}
  \textbf{\bibinfo{volume}{C72}}, \bibinfo{pages}{1878} (\bibinfo{year}{2012}),
  \eprint{1110.3568}.

\bibitem[{\citenamefont{Susskind}(1979)}]{Susskind:1978ms}
\bibinfo{author}{\bibfnamefont{L.}~\bibnamefont{Susskind}},
  \bibinfo{journal}{Phys.Rev.} \textbf{\bibinfo{volume}{D20}},
  \bibinfo{pages}{2619} (\bibinfo{year}{1979}).

\bibitem[{\citenamefont{Barbieri and Giudice}(1988)}]{Barbieri:1987fn}
\bibinfo{author}{\bibfnamefont{R.}~\bibnamefont{Barbieri}} \bibnamefont{and}
  \bibinfo{author}{\bibfnamefont{G.}~\bibnamefont{Giudice}},
  \bibinfo{journal}{Nucl.Phys.} \textbf{\bibinfo{volume}{B306}},
  \bibinfo{pages}{63} (\bibinfo{year}{1988}).

\bibitem[{\citenamefont{Ellis et~al.}(1986)\citenamefont{Ellis, Enqvist,
  Nanopoulos, and Zwirner}}]{Ellis:1986yg}
\bibinfo{author}{\bibfnamefont{J.~R.} \bibnamefont{Ellis}},
  \bibinfo{author}{\bibfnamefont{K.}~\bibnamefont{Enqvist}},
  \bibinfo{author}{\bibfnamefont{D.~V.} \bibnamefont{Nanopoulos}},
  \bibnamefont{and} \bibinfo{author}{\bibfnamefont{F.}~\bibnamefont{Zwirner}},
  \bibinfo{journal}{Mod.Phys.Lett.} \textbf{\bibinfo{volume}{A1}},
  \bibinfo{pages}{57} (\bibinfo{year}{1986}).

\bibitem[{\citenamefont{Kadastik et~al.}(2012)\citenamefont{Kadastik, Kannike,
  Racioppi, and Raidal}}]{Kadastik:2011aa}
\bibinfo{author}{\bibfnamefont{M.}~\bibnamefont{Kadastik}},
  \bibinfo{author}{\bibfnamefont{K.}~\bibnamefont{Kannike}},
  \bibinfo{author}{\bibfnamefont{A.}~\bibnamefont{Racioppi}}, \bibnamefont{and}
  \bibinfo{author}{\bibfnamefont{M.}~\bibnamefont{Raidal}},
  \bibinfo{journal}{JHEP} \textbf{\bibinfo{volume}{1205}}, \bibinfo{pages}{061}
  (\bibinfo{year}{2012}), \eprint{1112.3647}.

\bibitem[{\citenamefont{Fowlie et~al.}(2013)\citenamefont{Fowlie, Kowalska,
  Roszkowski, Sessolo, and Tsai}}]{Fowlie:2013oua}
\bibinfo{author}{\bibfnamefont{A.}~\bibnamefont{Fowlie}},
  \bibinfo{author}{\bibfnamefont{K.}~\bibnamefont{Kowalska}},
  \bibinfo{author}{\bibfnamefont{L.}~\bibnamefont{Roszkowski}},
  \bibinfo{author}{\bibfnamefont{E.~M.} \bibnamefont{Sessolo}},
  \bibnamefont{and} \bibinfo{author}{\bibfnamefont{Y.-L.~S.}
  \bibnamefont{Tsai}}, \bibinfo{journal}{Phys.Rev.}
  \textbf{\bibinfo{volume}{D88}}, \bibinfo{pages}{055012}
  (\bibinfo{year}{2013}), \eprint{1306.1567}.

\bibitem[{\citenamefont{Roszkowski et~al.}(2011)\citenamefont{Roszkowski,
  Ruiz~de Austri, Trotta, Tsai, and Varley}}]{Roszkowski:2009sm}
\bibinfo{author}{\bibfnamefont{L.}~\bibnamefont{Roszkowski}},
  \bibinfo{author}{\bibfnamefont{R.}~\bibnamefont{Ruiz~de Austri}},
  \bibinfo{author}{\bibfnamefont{R.}~\bibnamefont{Trotta}},
  \bibinfo{author}{\bibfnamefont{Y.-L.~S.} \bibnamefont{Tsai}},
  \bibnamefont{and} \bibinfo{author}{\bibfnamefont{T.~A.}
  \bibnamefont{Varley}}, \bibinfo{journal}{Phys.Rev.}
  \textbf{\bibinfo{volume}{D83}}, \bibinfo{pages}{015014}
  (\bibinfo{year}{2011}), \eprint{0903.1279}.

\bibitem[{\citenamefont{Profumo}(2013)}]{Profumo:2013yn}
\bibinfo{author}{\bibfnamefont{S.}~\bibnamefont{Profumo}}
  (\bibinfo{year}{2013}), \eprint{1301.0952}.

\bibitem[{\citenamefont{Heinemeyer et~al.}(2013)\citenamefont{Heinemeyer,
  Hollik, Weiglein, and Zeune}}]{Heinemeyer:2013dia}
\bibinfo{author}{\bibfnamefont{S.}~\bibnamefont{Heinemeyer}},
  \bibinfo{author}{\bibfnamefont{W.}~\bibnamefont{Hollik}},
  \bibinfo{author}{\bibfnamefont{G.}~\bibnamefont{Weiglein}}, \bibnamefont{and}
  \bibinfo{author}{\bibfnamefont{L.}~\bibnamefont{Zeune}},
  \bibinfo{journal}{JHEP} \textbf{\bibinfo{volume}{1312}}, \bibinfo{pages}{084}
  (\bibinfo{year}{2013}), \eprint{1311.1663}.

\bibitem[{\citenamefont{Alves et~al.}(2012)}]{Alves:2011wf}
\bibinfo{author}{\bibfnamefont{D.}~\bibnamefont{Alves}} \bibnamefont{et~al.}
  (\bibinfo{collaboration}{LHC New Physics Working Group}),
  \bibinfo{journal}{J.Phys.} \textbf{\bibinfo{volume}{G39}},
  \bibinfo{pages}{105005} (\bibinfo{year}{2012}), \eprint{1105.2838}.

\bibitem[{\citenamefont{Cohen et~al.}(2013)\citenamefont{Cohen, Golling, Hance,
  Henrichs, Howe et~al.}}]{Cohen:2013xda}
\bibinfo{author}{\bibfnamefont{T.}~\bibnamefont{Cohen}},
  \bibinfo{author}{\bibfnamefont{T.}~\bibnamefont{Golling}},
  \bibinfo{author}{\bibfnamefont{M.}~\bibnamefont{Hance}},
  \bibinfo{author}{\bibfnamefont{A.}~\bibnamefont{Henrichs}},
  \bibinfo{author}{\bibfnamefont{K.}~\bibnamefont{Howe}}, \bibnamefont{et~al.}
  (\bibinfo{year}{2013}), \eprint{1311.6480}.

\end{thebibliography}
\end{document}